\documentclass{emulateapj}

\journalinfo{Astrophysical Journal}
\slugcomment{Received 09 Oct 2007; Accepted 03 Dec 2007}

\def\p{\partial}

\def\nab{\mbox{\boldmath $\nabla$}}

\def\rb{\bar{\rho}}
\def\tb{\bar{T}}
\def\sb{\bar{S}}

\def\vph{\hat{v}_{\phi}}

\def\vvr{\tilde{v}}

\def\bbr{\tilde{B}}

\newcommand{\cross}{\mbox{\boldmath $\times$}}

\begin{document}

\title{Simulations of dynamo action in fully convective stars}

\author{Matthew K. Browning \altaffilmark{1}}

\affil{Astronomy Dept, 601 Campbell Hall, UC Berkeley, Berkeley CA
  94720-3411, matthew@astro.berkeley.edu}

\altaffiltext{1}{present address: Dept of Astronomy \& Astrophysics, U. Chicago,
  5640 S. Ellis Ave, Chicago, IL 60637}

\begin{abstract}

  We present three-dimensional nonlinear magnetohydrodynamic simulations of
  the interiors of fully convective M-dwarfs.  Our models consider 0.3
  solar-mass stars using the Anelastic Spherical Harmonic code, with the
  spherical computational domain extending from 0.08-0.96 times the overall
  stellar radius.  Like previous authors, we find that fully convective
  stars can generate kG-strength magnetic fields (in rough equipartition
  with the convective flows) without the aid of a tachocline of shear.
  Although our model stars are everywhere unstably stratified, the
  amplitudes and typical pattern sizes of the convective flows vary
  strongly with radius, with the outer regions of the stars hosting
  vigorous convection and field amplification while the deep interiors are
  more quiescent.  Modest differential rotation is established in
  hydrodynamic calculations, but -- unlike in some prior work --
  strongly quenched in MHD simulations because of the Maxwell stresses
  exerted by the dynamo-generated magnetic fields.  Despite the lack of
  strong differential rotation, the magnetic fields realized in the
  simulations possess significant mean (axisymmetric) components, which we
  attribute partly to the strong influence of rotation upon the slowly
  overturning flows.

\end{abstract}


\section{INTRODUCTION}

Magnetic fields are ubiquitous in stars.  Across a wide range of stellar
masses, these fields are thought to arise from dynamo action, with
convection, rotation, and shear all likely players in the process of field
generation (e.g., Moffatt 1978).  Stars whose interiors are everywhere
convectively unstable have been thought to harbor dynamos that may differ
fundamentally from those in stars that also possess radiative cores (e.g.,
Durney, De Young \& Roxburgh 1993).  Recent observations and theoretical
models are greatly complicating this basic view, and a thorough
understanding of the dynamo process in such stars remains elusive.  We
begin here by outlining both the current theoretical understanding of the
magnetism of stars with and without internal stable zones, as well as some
recent observational puzzles that serve to motivate our work.  

\subsection{Magnetic fields in solar-type stars}

In the Sun, the boundary layer between the convective envelope and the
stably stratified core is believed to play a pivotal role in the generation
of global magnetic fields by dynamo action (e.g., Ossendrijver 2003).
Helioseismology has revealed that this interface region is a site of strong
shear, where the solar angular velocity profile transitions from
differential rotation -- with a fast equator, slow poles, and angular
velocity nearly constant on radial lines within the convection zone -- to
solid-body rotation within the radiative core (e.g., Thompson et al. 1996).
In the standard ``interface dynamo'' paradigm for the global solar dynamo,
this shearing layer, called the tachocline, stretches and amplifies
poloidal poloidal magnetic fields generated within the convection zone,
giving rise to organized toroidal field structures (e.g., Parker 1993;
Charbonneau \& MacGregor 1997; Ossendrijver 2003).  These toroidal fields
may ultimately become unstable to magnetic buoyancy instabilities and rise
through the convection zone, with some eventually appearing at the surface
as sunspots and others being shredded by the convection and used to create
poloidal field, thereby completing the dynamo cycle.  Alternatively,
it has been argued that coherent meridional circulations may bring fields
to the surface (e.g., Rempel 2006; Dikpati \& Charbonneau 1999).  In either
case, the tachocline is likely a key element in the solar dynamo process --
partly because it is a site of strong shear, but also because the stable
stratification that prevails below the convection zone allows fields to be
greatly amplified before they become unstable to magnetic buoyancy
instabilities and rise.  In the convection zone, by contrast, the timescale
for field amplification to the $\sim$2000-3000 G commonly observed in sunspots
(e.g., Simon et al. 1988; Stix 2002) is longer than simple estimates of the timescale
for flux tubes to rise due to this instability (Parker 1975).

Simulations and mean-field models have helped to affirm the likely
importance of the tachocline in generating the large-scale magnetism of
stars like the Sun.  Three-dimensional magnetohydrodynamic (MHD)
simulations that modeled convection zones in various geometries have
demonstrated that helical convection can readily build strong (kG) magnetic
fields; in some cases (e.g., Jones \& Roberts 2000; Stellmach \& Hansen
2004), both small and large-scale fields were realized.  However, several
recent calculations of turbulent flows (at finite Prandtl number) have
suggested that generating a large-scale field may be quite difficult unless
the flow is highly organized (see discussion in Cattaneo \& Hughes 2006);
other authors have argued that boundary conditions may play a crucial role
in allowing large-scale field generation (e.g., Blackman \& Field 2000).
Recent 3-D spherical shell simulations that modeled the bulk of the solar
convective envelope (Brun, Miesch \& Toomre 2004, hereafter BMT04) have lent
some support to the view that convection alone may have difficulty in
building solar-like magnetism: dynamo action was realized, but the fields
tended to be mostly on small spatial scales, exhibited no evident parity
preferences, and showed a tendency to reverse in polarity on irregular
intervals of only a few hundred days.  Building upon this work, recent
simulations also included penetration by the turbulent convection into a
stable region with an imposed tachocline of shear (Browning et al. 2006).
These calculations showed that large-scale toroidal fields could indeed be
realized within the tachocline; further, these fields possessed
antisymmetric parity like that observed in sunspots, and showed much more
stable polarity than in simulations of the convection zone alone.
Likewise, extensive mean-field modeling has generally also suggested that
differential rotation in the tachocline plays a dominant role in generating
toroidal fields from poloidal fields, a process parameterized as the
$\Omega$-effect (e.g., Moffatt 1978; Steenbeck, Krause \& Radler 1966).  In
the Sun, this is taken to act in concert with the $\alpha$-effect that can
produce toroidal field from poloidal (and vice versa), either through the
action of cyclonic convection or through Babcock-Leighton effects near the
surface (e.g., Parker 1955; Steenbeck et al 1966; Babcock 1961; Leighton
1969).  Together, these constitute the solar $\alpha-\Omega$ dynamo.

\subsection{Puzzles of M-star magnetism}

Stars less massive than about 0.35 solar masses are fully convective, and
so cannot possess a transition region precisely like the solar tachocline.
It therefore seems natural to expect that they should harbor
qualititatively different magnetic dynamo action than stars like the Sun
(e.g., Durney, De Young \& Roxburgh 1993).  Yet no obvious transition in
magnetic activity has been observed at spectral types $\sim$ M3, where the
stably stratified core disappears.  Instead, stars on both sides of this
``tachocline divide'' appear to be able to build magnetic fields
effectively.  Many fully convective stars are observed to have strong
chromospheric H$\alpha$ emission (e.g., Hawley, Gizis \& Reid 1996; Mohanty
\& Basri 2003; West et al. 2004), which is well-established as an
indication of the presence of magnetic fields.  Indeed, the fraction of
stars that show such emission increases markedly after the transition to
full convection, reaching a maximum around spectral type M8 (West et
al. 2004).  Magnetic fields have also recently been directly detected on
fully convective stars using magnetically sensitive FeH line ratios
(Reiners \& Basri 2007).

It remains unclear whether the magnetic fields in such stars differ
fundamentally -- either in spatial structure, temporal variability, or
dependence on stellar parameters like rotation rate -- from those in more
massive stars.  In stars with spectral types ranging from mid-F to early M,
observations indicate that chromospheric and coronal activity increase
rapidly with increasing rotational velocity, then saturate above a
threshold velocity (e.g., Noyes et al. 1984; Pizzolatto et al 2003;
Delfosse et al. 1998).  This threshold velocity appears to lessen with
decreasing stellar mass.  Some evidence also exists for a
``super-saturation'' regime, with magnetic activity somewhat lessened in
the most rapid rotators (e.g., James et al. 2000).  The situation is less
clear in the mid to late M-dwarfs.  Mohanty \& Basri (2003) argued that a
sample of stars ranging from M4 to M9 exhibited a common ``saturation
type'' rotation-activity relationship, with observed activity roughly
independent of rotation rate above a threshold value.  Measuring the
rotation rates of the slowest rotators is difficult, so it remains unclear
whether magnetic activity in these stars increases gradually with rotation
as in solar-like stars, or instead changes more abruptly.  Very recently,
two fully convective M6 stars have been found with reasonably rapid
rotation (v sin i $>$ 5 km s$^{-1}$) but no measurable chromospheric
emission (West \& Basri 2007, in preparation): these stars are not so cool
that low conductivity effects are likely to play a major role (see Mohanty
et al. 2002), so these observations raise further questions about how the
magnetic fields of such stars are influenced by rotation.  Another striking
constraint on the strength and morphology of M-dwarf magnetic fields has
been provided by Donati et al. (2006), who used Zeeman Doppler imaging to
show that the rapidly rotating fully convective star v374 Peg possesses a
large-scale axisymmetric field of kG strength, but no surface differential
rotation.

An additional complication is that the rotation rates of stars (and hence
perhaps their magnetic activity) depend strongly upon their age: stars
generally arrive on the main sequence rapidly rotating, and slow over time
through angular momentum loss via a magnetized wind (e.g., Skumanich 1972;
Weber \& Davis 1967).  The amount of rotational braking a star undergoes at
any point in its evolution is then presumably dependent on the strength of
its magnetic field, and perhaps also on the geometry of that field.  Thus,
differences between the dynamo acting on either side of the ``tachocline
divide'' might manifest themselves as differences in the rotational
evolution of stars with or without radiative cores.  Indeed, some recent
evidence suggests that the timescale for magnetic braking, as indicated by
the typical ages at which stars are no longer observably rotating, may
increase markedly at approximately the mass where full convection sets in
(e.g., West et al. 2007; Reiners et al. 2007; Barnes 2003).

\subsection{Prior modeling and this work}

A few previous authors have examined the generation of magnetic fields in
fully convective stars using either semi-analytical theory or simulation,
but the overall picture that emerges from these investigations is somewhat
murky.  Durney, De Young \& Roxburgh (1993) argued that in the absence of a
tachocline of shear, the magnetic fields of fully convective stars should
be dominated by small-scale dynamo action, with typical spatial scales of
order the size of convective cells.  Kuker \& Rudiger (1999) examined
dynamos in fully convective pre-main-sequence stars using mean field
theory; they assumed that such stars rotate approximately as rigid bodies,
and adopted a simple $\alpha$-quenching formula to account crudely for the
back-reaction of dynamo-generated fields upon the flows.  They found that
$\alpha^2$ dynamo solutions could be excited for moderate rotation rates,
giving rise to steady, non-axisymmetric mean fields.  Chabrier \& Kuker
(2006) performed analogous mean-field modeling for fully-convective
low-mass stars, typically also assuming no differential rotation, and found
large-scale non-axisymmetric fields were generated by an $\alpha^2$ dynamo;
these fields were steady and symmetric with respect to the equatorial
plane.  Chabrier \& Kuker (2006) also constructed a model with internal
differential rotation (which they argued might apply to brown dwarfs with
conductive cores), and found that predominantly toroidal axisymmetric
fields were generated by the $\alpha^2-\Omega$ dynamo that resulted.
Finally, Dobler, Stix \& Brandenburg (2006; hereafter DSB06) conducted 3-D
hydrodynamic and MHD simulations of fully-convective spheres using a
Cartesian grid-based finite-difference code.  They found that such stars
established ``anti-solar'' differential rotation, with the poles rotating
more rapidly than the equator.  Dynamo action was realized, ultimately
yielding typical magnetic field strengths approximately in equipartition
with the flows near the surface.  The resulting magnetic fields possessed
structure on a range of spatial scales, with a substantial large-scale
component.

The origins of the discrepancies between these theoretical predictions are
unclear.  Part of the difficulty lies in the inherent limitations of
mean-field modeling, which separates the flows and fields into a
large-scale (mean) component and everything else, with the latter typically
parameterized in terms of the mean fields using a turbulence closure
model.  Such modeling cannot provide detailed descriptions of the
spatial distribution of the magnetic fields; nor can it independently
constrain the field strengths that are achieved by dynamo action, since the
models usually adopt $\alpha$-quenching prescriptions that simply eliminate
the $\alpha$-effect as equipartition is reached.  Further, the differential
rotation realized by the convection was essentially a free parameter within
the models quoted above.  Nonetheless, the results of such modeling are
highly suggestive of the roles played by convection and rotation in
building magnetic fields on both large and small scales.  Indeed, a variety
of related dynamo-theoretic work, in particular calculations of MHD spectra
within the eddy-damped quasi-normal Markovian approximation (Pouquet
et al. 1976) indicates that convection with helicity -- as imparted by
rotation -- may lead to cascades of magnetic energy from small toward large
scales, again hinting that shear need not be present in order to generate
large-scale magnetic fields.  Meanwhile, the results of the numerical
simulations thus far cannot be taken as the last word on the subject
either: although they provide descriptions of the dynamics on many
different scales, they are still limited by numerical resolution to
parameter regimes far removed from those of stellar convection.  Further,
it is difficult to gauge {\sl a priori} how the many different
simplifications adopted within DSB06 (or any other simulation) might impact
their results.  In any event, the conflict between the results of all
theoretical models published so far and the observational constraints on
field geometry provided by Donati et al. (2006) lend further vibrancy to
the study of magnetism in fully convective stars.  

Motivated by the wealth of observational and theoretical puzzles posed by
such stars, we turn in this paper to new 3-D MHD simulations of the
interiors of low-mass M-dwarfs.  We aim here to provide additional
constraints on the nature of convection and differential rotation in these
stars, on the possible dynamo action achieved within their interiors, and
on the morphology of the magnetic fields generated by such dynamo action.
Our work is most analogous to that of DSB06, but we differ from them in
several ways -- both in the construction of our model and in the
results it produces.  In \S 2 below, we describe our numerical model and
the principal simplifications adopted, highlighting some of the differences
between our simulations and those of DSB06.  Section 3 contains a
description of the morphology of the convective flows, and an analysis of
the spatial variations present in those flows.  In \S 4, we assess the
overall energetics of the dynamo action that is realized, and in \S 5 we
describe the morphology, strength, and temporal variability of the
resulting magnetism.  The establishment of differential rotation in
hydrodynamic cases, and its quenching in MHD ones, is analyzed in \S 6.  We
assess some aspects of the dynamo process itself in \S 7; we close in \S 8
with a comparison to prior modeling and observation, and a reflection on
what remains to be done.  

\section{FORMULATING THE PROBLEM}

\subsection{Fully convective rotating sphere}

The simulations here are intended to be highly simplified descriptions of
the interiors of fully convective 0.3 solar mass M-dwarfs.  We utilize the
Anelastic Spherical Harmonic (ASH) code, which solves within the anelastic
approximation the 3-D equations that govern fluid motion and magnetic field
evolution.  Our computational domain is spherical, and extends from 0.08R
to 0.96R in radius, with R the overall stellar radius of $2.068 \times
10^{10}$ cm.  We are forced to exclude the inner 8\% of the star from our
computations, both because the coordinate systems employed in ASH are
singular there, and because the small numerical mesh sizes very near the
center of the star would necessitate impractically small timesteps.  This
excluded central region might in principle cause some spurious physical
behavior, for instance by projecting a Taylor column aligned with the
rotation axis (e.g., Pedlosky 1987) into the surrounding fluid, by giving
rise to a central viscous boundary layer, or simply by excluding motions
that would otherwise pass through the stellar center.  In trial simulations
with smaller and larger excluded central regions (ranging from 0.04R to
0.12R), the properties of the mean flows were very similar to those
described here, giving us some confidence that the large-scale dynamics are
relatively insensitive to this inner boundary layer.  Our computations also
do not extend all the way to the stellar surface, because the very low
densities in the outer few percent of the star favor the driving of fast,
small-scale motions that we cannot resolve.  

The initial stratifications of mean density $\bar{\rho}$, energy generation
rate $\epsilon$, gravity $g$, radiative diffusivity $\kappa_{\rm rad}$, and
entropy gradient $dS/dr$ are adopted from a 1-D stellar model (I. Baraffe,
private communication; cf Chabrier et al. 2000).  The initial
profiles of the remaining thermodynamic quantities are then given by
solving the system of equations described in \S2.2, as described more
thoroughly elsewhere (Miesch et al. 2000; BMT04).
The thermodynamic quantities are updated throughout the course of the
simulation, as the evolving convection modifies the spherically symmetric
mean state.

The main parameters of our hydrodynamic and magnetohydrodynamic (MHD)
simulations are described in Table 1.  The three hydrodynamic cases $A$,
$B$, and $C$ sample flows of varying complexity, achieved by modifying the
effective eddy viscosities and diffusivities $\nu$ and $\kappa$, with case
$C$ the most turbulent (and highest-resolution) simulation.  Two MHD
calculations Cm and Bm were begun by introducing small-amplitude seed
magnetic fields into the statistically mature progenitor cases C and B
respectively, and allowing the fields to evolve.  A third MHD case Cm2 was
begun from a statistically mature instant in the evolution of case Cm. The
two cases Cm and Cm2 differ in the magnetic Prandtl number $Pm=\nu/\eta$
used, which in turn affects the magnetic Reynolds numbers $Rm \sim uL/\eta$
achieved by the evolved simulations.  All the simulations presented here
rotate at the solar angular velocity of $\hat{\Omega}=2.6 \times 10^{-6}$
s$^{-1}$.  In this paper, we have chosen for clarity's sake to concentrate
our discussion on the two cases C and Cm.  These represent our highest $Re
\sim uL/\nu$ and $Rm$ cases, respectively, so we believe they are likely to
be most indicative of the highly turbulent conditions achieved in real
stars.  Where appropriate, we give some indications of the behavior of the
other cases.

\begin{deluxetable*}{ccccccc}
\tablecolumns{7}
\tablenum{1}
\tablecaption{Simulation Attributes}
\tablehead{
\colhead{Case} & \colhead{{\sl A}} & \colhead{{\sl B}} & \colhead{{\sl C}}
& \colhead{{\sl Bm}} & \colhead{{\sl Cm}}  & \colhead{{\sl Cm2}}}
\startdata
  &   &  Input parameters &  & \\
  \hline

  $\nu$ (cm$^2$ s$^{-1}$) & $5.0 \times 10^{11}$ & $2.2 \times 10^{11}$ &
  $1.0 \times 10^{11}$ & $2.2 \times 10^{11}$ & $1.0 \times 10^{11}$ & $1.0
  \times 10^{11}$ \\
  $\kappa$ (cm$^2$ s$^{-1}$) & $2.0 \times 10^{12}$ & $8.8 \times 10^{11}$
  & $4.0 \times 10^{11}$ &  $8.8 \times 10^{11}$ & $4.0 \times 10^{11}$ &
  $4.0 \times 10^{11}$ \\
  $T_a$ & 1.1 $\times 10^7$ & 5.9 $\times 10^7$ & 2.8 $\times 10^8$ & 5.9
  $\times 10^7$ & 2.8 $\times 10^8$ & 2.8 $\times 10^8$ \\
  $\eta$ (cm$^2$ s$^{-1}$) & -- & --
  & -- &  $2.75 \times 10^{10}$ & $2.0 \times 10^{10}$ &
  $1.25 \times 10^{10}$ \\  
  $P_m$ & -- & -- & -- & 8 & 8 & 5 \\
  \hline
  &  & Measured quantities &  & \\
  \hline
  $R_a$ & 3.4 $\times 10^5$ & 1.4 $\times 10^6$ & 6.5 $\times 10^6$ & 1.7
  $\times 10^6$ & 6.3 $\times 10^6$ & 6.0 $\times 10^6$ \\
  $R_e$  & 65 & 120 & 270 & 110 & 210 & 230 \\
  $R_m$  &  -- & -- & -- & 880 & 1650 & 1160  \\
  $R_o$ & 1.6 $\times 10^{-2}$ & 1.3 $\times 10^{-2}$ & 1.3 $\times 10^{-2}$ & 1.2
  $\times 10^{-2}$ & 1.0 $\times 10^{-2}$ & 1.1 $\times 10^{-2}$ \\  
  $R_c$ & 0.38 & 0.34 & 0.33 & 0.37 & 0.33 & 0.32 \\  

\enddata
\tablecomments{The Prandtl number $P_r=\nu/\kappa=0.25$ for all
simulations; the magnetic Prandtl number $P_m=\nu/\eta$ is indicated for
each case, along with the viscosity $\nu$, eddy thermal diffusivity
$\kappa$, and magnetic diffusivity $\eta$ (in cm$^{2}$s$^{-1}$).  The
rotation rate $\Omega=2.6 \times 10^{-6}$ s$^{-1}$ for all cases.
Evaluated as volume averages are the Rayleigh number $R_a=(-\p
\bar{\rho}/\p S)\Delta S g L^3/\rho \nu \kappa$ (with $\Delta S$ the entropy
contrast across the interior),  the Taylor number $T_a = 4 \Omega^2
L^4/\nu^2$, and the convective Rossby number $R_c=\sqrt{R_a/T_a P_r}$.  The Reynolds number
$R_e=\vvr' L/\nu$, the magnetic Reynolds number $R_m=\vvr' L/\eta$, and the
Rossby number $R_o = \vvr'/2 \Omega L$ are evaluated at $r=0.88R$ using the
rms velocity $\vvr$ there.  Values based on the maximum velocity would be
about a factor of four higher; likewise, values would be slightly higher if
based on $\vvr$ closer to the surface, and lower if based on $\vvr$ deeper
within the star.  }
\end{deluxetable*}

\subsection{Anelastic MHD Equations}

ASH solves the 3-D MHD anelastic equations of motion in a rotating
spherical geometry using a pseudospectral semi-implicit approach (e.g.,
Clune et al. 1999; Miesch et al. 2000; BMT04).  The
equations are fully nonlinear in the velocities and magnetic fields, but
linearized in thermodynamic variables with respect to a spherically
symmetric mean state that is also allowed to evolve.  This mean state is
taken to have density $\bar{\rho}$, pressure $\bar{P}$, temperature
$\bar{T}$, specific entropy $\bar{S}$; perturbations are denoted as $\rho$,
$P$, $T$, and $S$.  The equations solved are
\begin{eqnarray}
\nab\cdot(\rb {\bf v}) &=& 0, \\
\nab\cdot {\bf B} &=& 0, \\
\rb \left(\frac{\p {\bf v}}{\p t}+({\bf v}\cdot\nab){\bf v}+2{\bf \Omega_o}\times{\bf v}\right) 
 &=& -\nab P + \rho {\bf g} \nonumber \\
+ \frac{1}{4\pi} (\nab\times{\bf B})\times{\bf
    B} 
&-& \nab\cdot\mbox{\boldmath $\cal D$}-[\nab\bar{P}-\rb{\bf g}], 
\\
\rb \tb \frac{\p S}{\p t}+\rb \tb{\bf v}\cdot\nab (\sb+S)&=& 
\nab\cdot[\kappa_r \rb c_p \nab (\tb+T) \nonumber \\
+\kappa \rb \tb \nab (\sb+S)] &+&\frac{4\pi\eta}{c^2}{\bf j}^2\nonumber \\
+2\rb\nu[e_{ij}e_{ij} &-& 1/3(\nab\cdot{\bf v})^2]
 + \rb {\epsilon},\\
\frac{\p {\bf B}}{\p t}=\nab\times({\bf v}\times{\bf B})&-&\nab\times(\eta\nab\times{\bf B}),
\end{eqnarray}
where ${\bf g}$ is acceleration due to gravity, ${\bf v}=(v_r,v_{\theta},v_{\phi})$ is the velocity in spherical coordinates in 
the frame rotating at constant angular velocity ${\bf \Omega_o}$, ${\bf B}=(B_r,B_{\theta},B_{\phi})$ is the magnetic field, 
${\bf j}=c/4\pi\, (\nab\times{\bf B})$ is the current density, $\kappa_r$ is the radiative diffusivity,
$c_p$ is the specific heat at constant pressure,  $\nu$ is the effective
eddy viscosity, $\kappa$ is the effective thermal diffusivity, $\eta$ is the 
effective magnetic diffusivity, and ${\bf \cal D}$ is the viscous stress
tensor, defined by
\begin{eqnarray}
{\cal D}_{ij}=-2\rb\nu[e_{ij}-1/3(\nab\cdot{\bf v})\delta_{ij}],
\end{eqnarray}
with $e_{ij}$ the strain rate tensor. The volume heating term $\rb \epsilon$ is included to
represent energy generation by nuclear burning of the CNO cycle within the
convective core, and is assumed for simplicity to scale with temperature
alone.  Our adoption of a sub-grid-scale
(SGS) heat transport term proportional to the entropy gradient in equation
(4) is most justifiable in convection zones, where the stratification will
tend toward adiabaticity; in stable zones, this term should be modified in
order to avoid spuriously large SGS heat fluxes directed radially inwards
(as discussed in Miesch 1998; Miesch et al. 2000).  In ASH, we choose to
deal with this difficulty by specifying the spherically symmetric
($\ell=0$) eddy thermal diffusivity $\kappa_0$ separately from the $\ell
\ne 0$ component $\kappa$; the latter is here taken to be constant in
radius, whereas the former increases in a narrow layer near the surface,
where the SGS transport must account for the entire outward energy flux
(see \S 3.2). The eddy diffusivity $\kappa$ is in effect purely
dissipative, and acts to smooth out entropy variations, whereas $\kappa_0$
is essentially a cooling term near the surface.  In simulations with stable
layers, $\kappa_0$ can be modified to account for the presence of radiative
regions (e.g., Miesch et al. 2000); here, however, this subgrid
transport is small throughout the interior because of the near-adiabatic
stratification.    To close the set of equations, we take the thermodynamic
fluctuations to satisfy the linearized relations
\begin{equation}\label{eos}
\frac{\rho}{\rb}=\frac{P}{\bar{P}}-\frac{T}{\tb}=\frac{P}{\gamma\bar{P}}
-\frac{S}{c_p},
\end{equation}
assuming the ideal gas law
\begin{equation}\label{eqn: gp}
\bar{P}={\cal R} \rb \tb ,
\end{equation}

\noindent where ${\cal R}$ is the gas constant.  The effects of
compressibility are included via the anelastic approximation, which filters out
sound waves that would otherwise limit the time steps allowed by the
simulation to the sound crossing time across the smallest computational
zone.  In the low Mach-number flows typical of stellar interior convection,
adopting the anelastic approximation allows us to take much larger
timesteps that satisfy the Courant-Freidrichs-Lewy condition imposed by the
convective flows, rather than the much shorter one required to follow
acoustic waves.  In the MHD simulations, the anelastic approximation
filters out fast magneto-acoustic modes but retains the Alfven and slow
magneto-acoustic modes.  We use a toroidal--poloidal decomposition for the
mass flux and magnetic field in order to ensure that both remain
divergence-free throughout the simulation, with
\begin{eqnarray}
{\rb\bf v}=\nab\times\nab\times (W {\bf e}_r) +  \nab\times (Z {\bf
  e}_r) , \\ 
{\bf B}=\nab\times\nab\times (C {\bf e}_r) +  \nab\times (A {\bf e}_r) ~~~, 
\end{eqnarray}
with ${\bf e}$ a unit vector, and involving the streamfunctions $W$,
$Z$ and magnetic potentials $C$, $A$.

This system of equations requires 12 boundary conditions and suitable
initial conditions. Because one of our aims is to assess the angular
momentum redistribution in our simulations, we have opted for torque-free
velocity boundary conditions at the top and bottom of the deep spherical
domain.  We have assumed

 $a$. impenetrable top and bottom surfaces: $v_r=0|_{r=r_{bot},r_{top}}$,

 $b$. stress free top and bottom: $\frac{\p}{\p
 r}\left(\frac{v_{\theta}}{r}\right)=\frac{\p}{\p
 r}\left(\frac{v_{\phi}}{r}\right)=0|_{r=r_{bot},r_{top}}$,

 $c$. constant entropy gradient at top and bottom: $\frac{\p \sb}{\p
r}=$ constant$|_{r=r_{bot},r_{top}}$,

 $d$. match to an external potential field at the top: ${\bf B} = \nabla
{\Phi} \rightarrow \Delta \Phi =0|_{r=r_{top}}$, and to a perfect
conductor (purely tangential field) at the bottom.

In the analysis that follows, we will often form various spatial and
temporal averages of the evolving convective flows and magnetic fields.
For clarity, we note that we use the symbol $\hat{a}$ to indicate temporal
and longitudinal averaging of a variable $a$, and the symbol $<a>$ to
denote longitudinal averaging to obtain the axisymmetric component of the
variable. Such averaging allows us to separate the fluctuating (denoted by
a prime as $a'$) from the axisymmetric (mean) parts of the variable.  The
symbol $\tilde{a}$ designates the rms average of $a$, carried out on a
spherical surface for many realizations in time.  Likewise, the combined
symbols $\tilde{a}'$ represent rms averaging of the variable from which the
axisymmetric portion has been subtracted.

\subsection{Numerical Approach}

No numerical simulation can model with perfect fidelity the intensely
turbulent convection occurring in stars.  The range of spatial scales
present in such convection is too vast; some simplification is
unavoidable. We choose in our global modeling to resolve the largest scales
of motion, which we believe are likely to play dominant roles in
redistributing energy and angular momentum and in building magnetic fields.
Our simulations are therefore classified as Large Eddy Simulations (LES),
with the effects of unresolved small scales of turbulence incorporated
using a sub-grid-scale (SGS) treatment.  Here these unresolved scales are
treated simply as enhanced viscosities and thermal and magnetic
diffusivities ($\nu$, $\kappa$, and $\eta$ respectively), which are thus
effective eddy viscosities and diffusivities.  For simplicity, we have
taken these to be constant in radius.  This implies that the viscous
damping at depth in our simulations may be too severe relative to that near
the surface.  More sophisticated SGS treatments, in which $\nu$ and
$\kappa$ are proportional to properties of the resolved flow field (e.g.,
the velocity or strain rate) would certainly be desirable, and we hope to
explore such strategies in future work.  Our simulations are characterized
by nondimensional numbers relating these diffusivities to one another and
to the various terms in the momentum equation (3); some of these
dimensionless numbers necessarily take on very different values in our
modeling than in actual stellar interiors.  We have adopted a magnetic
Prandtl number $Pm = \nu/\eta > 1$ in our MHD cases, even though a $Pm$
based on microscopic viscosities and diffusivities would be much less than
unity.  The large $Pm$ here reflects unresolved turbulent mixing processes,
and allows us to achieve moderately high values of the magnetic Reynolds
number $Rm = uL/\eta$ with tractable numerical resolution.  At small $Pm$,
the critical $Rm$ needed for dynamo action increases considerably (Boldyrev
\& Cattaneo 2004; Schekochihin et al. 2005), rendering simulations more
computationally demanding.  The strength and morphology of the magnetic
fields realized in our simulations are likely sensitive at some level to
the $Pm$ and $Rm$ chosen. We are somewhat encouraged by prior simulations
of convection in the solar interior in comparable parameter regimes (Brun
\& Toomre 2002; Brun, Browning \& Toomre 2005; Browning et al. 2006), for
these have been relatively successful in matching the detailed
observational constraints on, e.g., solar differential rotation provided by
helioseismology.

The numerical strategies employed within ASH are described in detail
elsewhere (Clune et al. 1999; BMT04); we here summarize only a
few key features.  The dynamical variables within ASH are expanded in terms
of spherical harmonic basis functions $Y_l^m(\theta, \phi)$ in the
horizontal directions and Chebyshev polynomials $T_n(r)$ in the radial.
The spatial resolution is thus kept uniform everywhere on a spherical
surface by employing a complete set of spherical harmonics of degree
$\ell$, retaining all azimuthal orders $m$ in what is referred to as a
triangular truncation.  We limit our expansion to a degree $\ell=\ell_{\rm
  max}$, related to the number of latitudinal mesh points $N_{\theta}$ by
$\ell_{\rm max} = (2 N_{\theta} - 1)/3$, take $N_{\phi} = 2 N_{\theta}$
longitudinal mesh points, and use $N_r$ colocation points for the radial
projection onto Chebyshev polynomials.  Our highest resolution simulations
here (Cm) have $\ell_{\rm max} =340$ (implying $N_{\theta}=512$ and
$N_{\phi}=1024$) and $N_r=192$.  We employ an implicit, second-order
Crank-Nicholson scheme for calculating the time evolution of the linear
terms in equations (1-5), whereas an explicit second-order Adams-Bashforth
scheme is used for the advective, Lorentz, and Coriolis terms.  ASH runs
efficiently on a variety of massively parallel supercomputers using the
message passing interface (MPI).  The code's performance scales reasonably
well up to about 1000 processors.  The simulations described here required
roughly half a million hours of computing time.

Our computational approach differs from that of DSB06 in a few important
ways.  Because theirs is the only prior 3-D MHD simulation of fully
convective stars, we here comment briefly on these differences.  In order
to keep thermal relaxation timescales small, the simulations in DSB06
adopted a stellar luminosity roughly $10^{12}$ times higher than
appropriate for an actual M dwarf.  This choice is related to the fact
that DSB06 adopt the same thermal diffusivities for the mean temperature
gradient and for the small-scale turbulent temperature fluctuations; in our
modeling, the mean temperature gradient is acted on by the thermal
diffusivity $\kappa_r$ taken from a 1--D stellar model, whereas $\kappa$
for the turbulent temperature field is (as in DSB06) a sub-grid-scale eddy
diffusivity.  Our strategy allows us to assess with reasonable fidelity the
radiative flux within the interior, since $\kappa_r$ in our models is
ultimately set by the radiative opacities of the 1--D stellar model.  The
DSB06 rescaling of the luminosity implies a commensurate artificial
increase in the typical convective velocities.  In order to keep the ratio
between the convective turnover times and the rotation period approximately
correct, they also considered very rapid rotation rates.  In terms of the
nondimensional numbers $Re$, $Rm$, and $Ro$, our simulations are roughly
comparable to theirs.  However, it is difficult to gauge with certainty the
impact that their rescalings of $L$, $v_c$, and $\Omega$ may have on the
resulting flows and magnetic fields.  We have chosen not to perform such a
rescaling, which means that very long-term adjustments of the mean
temperature gradient would not be captured in our modeling.  Both
strategies for dealing with the thermal diffusivity have been widely
employed; neither is perfect.  Our simulations also differ from those of
DSB06 in a few smaller ways.  The overall density contrast between the
inner and outer boundaries in our models is about 100, consistent with the
contrast between 0.1 and 0.96R in the 1--D stellar model we used for our
initial conditions.  In DSB06, the density varied by a factor of about 5
from center to surface; the larger density contrasts in our modeling have
a substantial impact on the morphology of the convective flows.  The boundary
conditions adopted in DSB06 also differ from ours; theirs is closer to a
no-slip boundary condition than to the stress-free boundaries used here.
This difference may impact our results on the differential rotation
realized in hydrodynamic simulations (\S 6).  Because they employed a
Cartesian finite-difference code, DSB06 were able to model the central few
percent of the star, omitted in our modeling.  These factors may all
contribute to differences between the results here and in DSB06.
Nonetheless, there is some common ground between our findings and theirs;
we comment more on both the similarities and differences of the two models
in \S 8 below.

\begin{deluxetable*}{lllllll}
\tablecolumns{7}
\tablenum{2}
\tablecaption{Properties of Flows and Fields}
\tablehead{
\colhead{Case} & \colhead{A} & \colhead{B} & \colhead{C} & 
\colhead{Bm} & \colhead{Cm} & \colhead{Cm2}
}
\startdata
 KE & 6.4$\times 10^6$  & 1.0$\times 10^7$  & 2.6$\times 10^7$  &
 5.3$\times 10^6$ & 3.0 $\times 10^6$  & 3.8 $\times 10^6$   \\
 DRKE & 2.8$\times 10^6$  & 7.7$\times 10^6$  & 2.2$\times 10^7$  &
 2.8$\times 10^6$ & 4.8 $\times 10^5$  & 1.1 $\times 10^6$   \\
 CKE & 3.7$\times 10^6$  & 2.6$\times 10^6$  & 3.5$\times 10^6$  &
 2.5$\times 10^6$ & 2.5 $\times 10^6$  & 2.7 $\times 10^6$   \\ 
 ME/KE & -- & -- & --  & 50\% & 120\% & 90\% \\
 $\vvr'$(0.94R) & 24 & 22 & 23 & 19 & 19 & 19\\
 $\vvr'$(0.50R) & 4 & 4 & 4 & 4 & 4 & 4 \\
 $\bbr$(0.94R) & -- & -- & -- & 2000 & 6200 & 7200\\
 $\bbr$(0.50R) & -- & -- & -- & 6000 & 13100 & 10400\\
 $\Delta \Omega/\Omega$ & 8\% & 14\% & 22\% & 8\% & 2\% & 4\%\\
\enddata
\tablecomments{The kinetic energy density KE ($1/2 ~ \rb v^2$), averaged over volume
and time, is listed along with energy density of
the convection (CKE) and the differential rotation (DRKE), 
together with the average magnetic energy density ME
($B^2/8\pi$) (expressed, where appropriate, as a percentage of KE).  Also
indicated at two depths are the fluctuating rms velocity $\vvr'$ (m s$^{-1}$) and the
rms magnetic field strength (G).  Angular velocity contrast from equator to
60$\degr$ is indicated as percentage of overall frame rotation rate. }
\end{deluxetable*}

\section{CONVECTIVE FLOWS AND ENERGY TRANSPORT}

\subsection{Morphology of the Flows}

The convective flows realized in our simulations possess structure on many
spatial scales.  An instantaneous view of the flows in the hydrodynamic
case C is provided by Figure 1, which shows the radial velocity $v_r$ near
both the top of the computational domain (Fig. 1$a$, at $r=0.88R$) and
deeper within the interior (Fig. 1$b$, at r=0.24R).  At large radii, a
marked asymmetry between upflows and downflows is apparent (Fig. 1$a$),
with the downflows compact and strong while upflows are broader and weaker.
This asymmetry is driven mainly by the strong density stratification at
these radii: downflows tend to contract, whereas upflows expand.  Some of
these downflow lanes persist as coherent structures for extended intervals,
while many complex and intermittent features also appear on smaller scales.
Such coherent downflow plumes have previously been noted as a seemingly
generic feature of turbulent compressible convection (e.g., Brummell et
al. 2002; Brun \& Toomre 2002).

\begin{figure}[hpt]
  \center
  \epsscale{1.0}
  \includegraphics[width=3.4in, trim= 72 0 72 0]{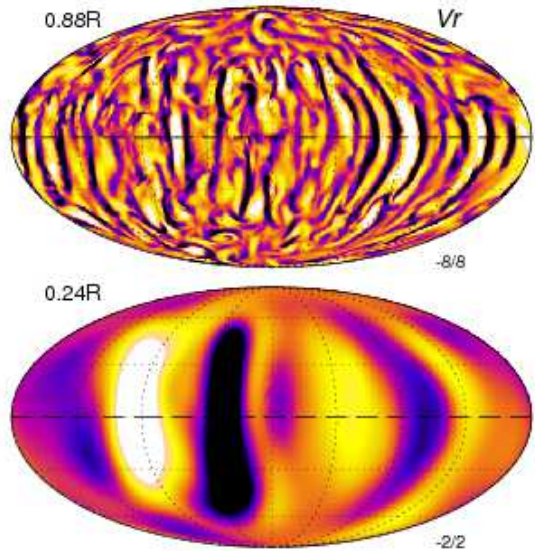}
  \caption{\label{vr2depths}Radial velocity $v_r$ on spherical surfaces at
    two depths for a single instant in the evolution of case Cm.  Upflows
    are rendered in red tones, and downflows in blue; the maxima and minima
  of the colormaps are indicated.  The flows are stronger and on smaller
  spatial scales near the surface than they are are depth.}
\end{figure}

In Figure 1$a$, these downflows generally appear aligned with the rotation
axis at low latitudes; at high latitudes the distribution of upflows and
downflows is more isotropic.  Similar distinctions between flows at high
latitudes versus those near the equator have often been realized in
simulations of the solar convective envelope (e.g., Miesch et al. 2000),
where the difference was identified with the ``tangent cylinder'' formed by
projecting the lower boundary at the equator onto the upper boundary.  In
those simulations, the inner boundary prohibited connectivity between the
northern and southern hemispheres for flows far from the equator, whereas
motions at low latitudes could readily couple both hemispheres.  Here, the
inner boundary is sufficiently deep to allow connectivity between
high-latitude flows as well, but motions that span both hemispheres are
still realized only near the equator.  This may arise because the effects
of the Coriolis forces, which ultimately drive turbulent alignment with the
rotation axis, still vary with distance from the rotation axis (see, e.g.,
Busse 1970); the low-latitude flows may also reflect a lingering preference
for the most unstable modes in a rotating spherical shell, which at these
rotation rates tend to be symmetric fluid rolls perpendicular to the
equator (e.g., Gilman 1976).

Deeper within the interior (Fig. 1$b$), the convection is characterized by
broader, weaker flows that span large fractions of a hemisphere.  Upflows
and downflows are fairly symmetric in appearance there, likely because the
density stratification at depth is weaker.  The density scale height
$\lambda_p = P/g \rho$ varies from about $10^9$ cm at $r=0.88R$ to $4
\times 10^9$ cm at $r=0.24R$, which corresponds roughly to the physical
size of the convective cells at the two radii; note that the spatial size
of the convective patterns, not just their angular size, varies with
radius.  The convective patterns at $r=0.24R$ are reminiscent of sectoral
spherical harmonics $Y^{\ell}_{m}$ with $\ell = m =3$; at other instants in
the evolution of case C, this identification is stronger.  The motions deep
in the interior are linked to those farther out: small downflow plumes at
large radii merge as they descend, and coalesce to form the broad downflows
seen in Figure 1$b$.  This coupling between depths is weakly discernible in Figure
1: the most striking downflow plumes at $r=0.88R$ generally occur in
regions where downflow persists at $r=0.24R$.

The amplitude of the convective motions also changes with depth.  Typical
rms velocities $\tilde{v}$ at $r=0.88R$ are about 12 m s$^{-1}$, whereas at
$r=0.24R$, $\tilde{v} \approx 2$ m s$^{-1}$.  This variation, together with
the smaller typical eddy sizes $l_{\rm eddy}$ near the surface, means that
the local convective overturning time $\tau_c \sim l_{\rm eddy}/\tilde{v}$
varies by a factor of about 20 across the domain.  For small-scale local
dynamo action, the characteristic timescale for field amplification is
roughly the convective turnover time (e.g., Childress \& Gilbert 1995);
thus we might expect that fields will be built more rapidly in the outer
layers of the star than at depth.  We will see in \S 5 that this is indeed
the case.

Thus, although the interior is unstably stratified at all depths, there are
still two conceptually distinct regions: one near the surface in which
convection is vigorous, possesses a variety of small-scale structure, and
might quickly amplify any seed magnetic fields, and another at depth where
the flows are more quiescent, with large-scale overturning motions that may
amplify fields somewhat more slowly.  In the following subsection, we
examine why the vigor of convection may vary with depth.

\subsection{Spatial Variation of Energy Transport}

Convection in stars is driven ultimately by the need to transport energy
outwards.  That energy arises primarily from nuclear-burning reactions
within the central regions of the star, so the total luminosity $L(r)$ that
must be carried outwards is an increasing function of radius, out to the
point where nuclear burning stops and $L(r)$ is equal to the surface
luminosity $L_{\rm *}$. In Figure 2, we assess for simulation C the radial
transport of energy by different physical processes, defined as

\begin{equation}
F_c+F_k+F_{\rm r}+F_u+F_v=\frac{L(r)}{4\pi r^2},
\end{equation}
with
\begin{eqnarray}
F_c&=&\rb\, c_p\, \overline{v_r T'} \,,\\
F_k&=&\frac{1}{2}\, \rb\, \overline{v^2 v_r} \,,\\
F_r&=& -\kappa_{r}\, \rb\, c_p\, \frac{d\tb}{d r} \,, \\
F_u&=& -\kappa\, \rb\, \tb\, \frac{d\sb}{d r} \,, \\
F_v&=& -\overline{{\bf v}\cdot {\bf \cal D}} \,
\end{eqnarray}
where the overbar denotes an average over spherical surfaces and in time,
$F_c$ is the enthalpy flux due to resolved convective flows, $F_k$ the
kinetic energy flux, $F_{\rm r}$ the radiative flux, $F_u$ the unresolved
eddy flux, and $F_v$ the viscous flux.  The unresolved eddy flux $F_u$ is
the enthalpy flux from subgrid-scale motions that we cannot resolve, which
in ASH takes the form of a thermal diffusion operating on the mean entropy
gradient.  In MHD simulations (Cm, Bm, Cm2), the Poynting flux $F_m =
\frac{c}{4\pi} \overline{E_{\theta}B_{\phi}-E_{\phi}B_{\theta}}$ also
contributes, but is small.  The viscous flux and the kinetic energy flux
are generally small compared to $F_c(r)$ and $F_r(r)$; the unresolved flux
becomes large near the surface, where it carries all the energy because the
radial velocity is forced to vanish at the upper boundary.

\begin{figure}[hpt]
  \center
  \epsscale{1.0}
  \includegraphics[width=3.4in, trim= 36 0 36 0]{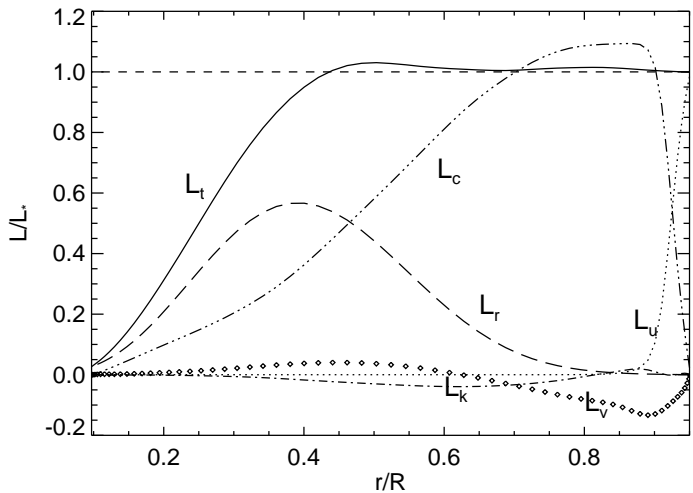}  
  \caption{\label{fluxbal}Time-averaged radial transport of energy in case
    C.  Shown are the convective enthalpy flux $F_c$, radiative flux $F_r$,
  viscous flux $F_v$, kinetic energy flux $F_k$, the unresolved flux $F_u$,
  and their sum the total flux $F_t$; all have been expressed as
  luminosities.  Although the stratification is convectively unstable
  everywhere, the convective flux carries most of the energy only at radii
  larger than about 0.45R.  }
\end{figure}

Figure 2 shows that the radiative flux carries most of the energy at small
radii, with $L_r \approx 0.70 L(r) \approx 0.10 L_{\rm *}$ at $r=0.15R$.
The enthalpy flux is smaller there, with $L_c \approx 0.30 L_r = 0.04
L_{\rm *}$.  Moving to larger radii, however, the total luminosity $L(r)$
rises up to $L_{\rm *}$ in accord with the continuing nuclear energy
generation, while at the same time the radiative luminosity $L_r(r)$
decreases.  Thus the energy that must be transported by convection goes up
significantly, rising to a maximum of $L_c \approx 1.1 L_{\rm *}$ at
$r=0.80R$; the excess over the stellar luminosity is mostly compensated by
an inward-directed viscous flux.  Although convection carries as little as
30\% of the local luminosity near the stellar center, the overall entropy
stratification is still nearly adiabatic.  The radiative flux is fixed by
the radiative opacity, here input from the 1-D stellar model, and by the
mean temperature gradient; likewise, the variation in the total luminosity
with radius is set by the nuclear energy generation rate $\epsilon (r)$,
also input from the 1-D model.  To the extent that these input properties
are accurate, and assuming there are no drastic long-term changes in the
prevailing temperature gradient (which we could not capture in our limited
simulation time), we believe that the significant variation of $L_c$ with
radius is likely to be a robust feature.  However, lower-mass stars
than those considered here might have smaller radiative fluxes in the deep
interior, owing to their lower central temperatures and hence higher
opacity from metals, and so may exhibit less radial variation of $L_c$.

This variation in the energy that must be transported by convection is
linked to the radial change in convective velocity noted in \S 3.1.  On
dimensional grounds, the typical convective velocity is given roughly by
$v_c \propto [L_c (R/M)]^{1/3}$, so the factor of 25 change in $L_c$ from
$r=0.15R$ to $r=0.88R$ would correspond to a factor of about 3 change in
typical convective velocity. This assumes that the correlations between
temperature fluctuations and the radial velocity field do not change
appreciably with depth; in practice, the energy transport is somewhat less
efficient amidst the highly turbulent flows near the surface, with many
regions where temperature and radial velocity are not as well-correlated as
they are at depth.  This effect, plus the strong variation of density with
radius, leads to even larger radial variations of $v_c$ than this simple
scaling argument would suggest.  

\section{DYNAMO ACTION REALIZED}

Magnetic dynamo action is achieved by the flows.  Small initial seed fields
are amplified by several orders of magnitude until they reach a
statistically equilibrated state in which their growth is balanced by Ohmic
decay.  The growth and saturation of the magnetic energy density in case Cm
is displayed in Figure 3$a$, while a phase of evolution after saturation is
examined in Figure 3$b$.  Also shown there are the kinetic energy densities
due to non-axisymmetric motions, which we term the convective kinetic
energy density (CKE) and the total kinetic energy density KE (which is the
sum of CKE, the energy in differential rotation DRKE, and the energy in
meridional circulations MCKE).  All are shown relative to the frame
rotating at $\Omega = 2.6 \times 10^{-6}$ s$^{-1}$. In the evolved
simulation, MCKE is approximately 300 times smaller than CKE, and DRKE is
likewise a factor of five smaller than CKE, so we have omitted them from
Figure 3 for clarity.

\begin{figure}[hpt]
  \center
  \epsscale{0.8}
  \includegraphics[width=3.4in, trim= 10 0 10 0]{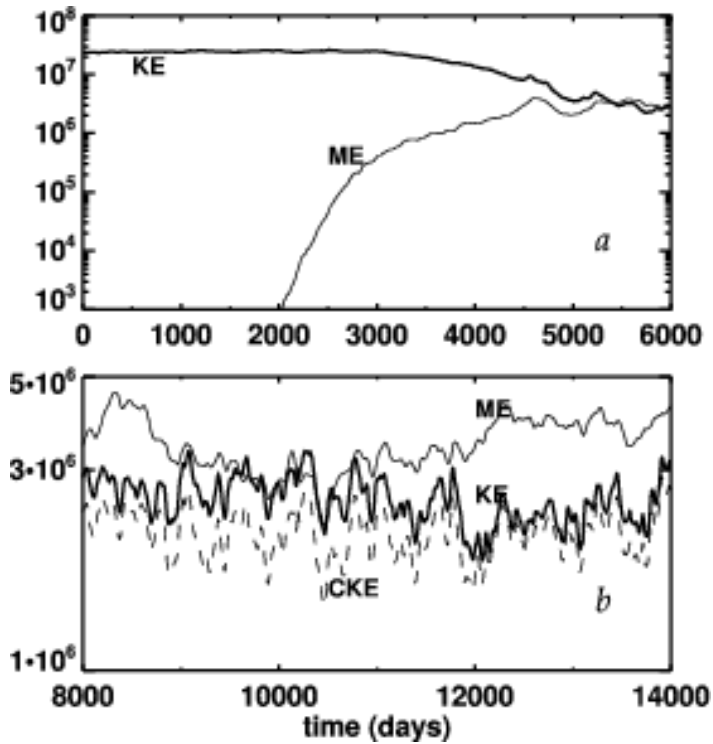}  
  \caption{\label{timetrace}Temporal evolution of the volume-averaged
    magnetic and kinetic energy densities in case Cm.  ($a$) The magnetic
    energy density (ME) grows by many orders of magnitude from its initial
    seed value, and ultimately equilibrates when comparable to the kinetic
    energy density (KE) relative to the rotating frame.  ($b$) Detailed
    view of the evolution of KE and ME during an interval after
    equilibration was reached; also shown is the energy density in the
    convection (CKE).  On average, the magnetic energy is about 120\% of KE
  and 140\% of CKE.}
\end{figure}

The magnetic energy in the simulations grows exponentially until it is
approximately in equipartition with the flows.  Over the last 200 days of
simulation C4m, a period during which no sustained growth or decay of the
various energy densities was evident, ME was approximately 120\%
of KE and about 140\% of CKE.  The initial seed value of ME was about
$10^{-2}$ ergs cm$^{-3}$, and the phase of exponential growth lasted about
2500 days, implying a characteristic e-folding timescale for magnetic
energy growth of about 150 days.  This may be compared to the typical eddy
convective turnover times in the simulation, which vary from about 20 days
near the surface to roughly 450 days at depth.

As the magnetic fields grow, they react back upon the flows that generated
them through the ${\bf j} \cross {\bf B}$ force term in equation (1).  The
net effect is to begin to reduce KE once ME reaches a threshold value of
about 5\% of KE; this is visible in Figure 3 starting at about 3000 days.
The reduction in KE is associated mainly with a large decline in DRKE,
whereas CKE appears largely unaffected by the growing fields.  In the
kinematic phase, DRKE is approximately 6 times CKE; after saturation of the
dynamo, DRKE/CKE is only about 0.2-0.4.  Although the energy densities
exhibit no systematic variation after the initial phase of dynamo growth,
they still show substantial short-term stochastic fluctuations.  During the
interval sampled by Figure 3, CKE varies by factors of about 3, and with it
the total KE; the magnetic energy density varies by similar amounts.  Thus
although ME $\approx$ 1.2 KE on average, it can rise as high as twice KE
for short intervals.

The other MHD simulations behave in a similar fashion.  In case Cm2, which
has a lower $Pm$ and hence lower $Rm$ (implying a less supercritical
dynamo), ME equilibrates at about 90\% of KE (120\% of CKE).  The weaker
magnetic fields in that case lead to a slightly smaller reduction in DRKE
than is realized in case Cm.  Case Bm, which is both more laminar (lower
$Re$) and less supercritical (lower $Rm$) than the other two simulations,
has still lower magnetic energy densities, with ME $\approx$ 50\% of KE.
These lower magnetic energy densities imply even less quenching of DRKE,
which varies between about 0.8-1.1 times CKE.  

It is instructive to compare these energy densities to those realized in
other convective dynamo simulations.  In numerical models of the solar
convective envelope (with $Pm=5$ and $R_m \approx 490$), Brun et al. (2004)
found that ME equilibrated at about 7\% of KE; Browning et al. (2006) found
comparable values within the convective envelope in simulations of the
convection zone and a forced tachocline, there adopting $Pm=8$. In modeling
fully convective stars with the PENCIL code, DSB06 found ME comparable to
KE in their most rapidly rotating runs.  Models of core convection in
A-type stars (Brun, Browning \& Toomre 2005, hereafter BBT05) yielded ME/KE typically
between 0.28 and 0.90, depending upon rotation rate and other simulation
parameters.  In those models, DRKE was strongly suppressed whenever ME was
greater than about 30\% of KE, leading to cyclical growth and decay of ME
and DRKE: large differential rotation led to growing ME, but the resulting
strong fields tended to suppress DRKE, which in turn resulted in a decline
in ME.  No cyclical behavior of this sort is observed in case Cm; instead,
ME is always much greater than DRKE.  The more laminar case Bm, with a
lower ME/KE ratio, does show some  linked oscillations in DRKE and ME,
suggesting that such intertwined feedback is realized only for a fairly
narrow range of magnetic energy densities.  We believe that case Cm is
likely to be more representative of the behavior of actual stellar
interiors at this rotation rate: if anything, the extraordinariliy high
$Re$ and $Rm$ realized in stars might be expected to lead to somewhat
higher magnetic energy densities than we find here, and hence perhaps to an
even stronger decline in DRKE.


\section{Properties of the Magnetic Fields}

\subsection{Morphology and Spatial Distribution}

The magnetic fields realized here possess both intricate small-scale
features as well as global-scale structures.  Like that of the flows that
sustain them, the morphology of the fields varies with radius, with the
typical length scale of the field increasing with depth.  A sampling of
such behavior is provided by Figure 4, which shows an instantaneous view of
the radial magnetic field $B_{r}$, the azimuthal magnetic field $B_{\phi}$,
and the radial velocity $v_r$ on spherical surfaces at two depths in case Cm.  Near
the surface, complex structures on many different scales continually emerge
and evolve, ranging from localized ripples to large-scale patches in $B_{\phi}$
that extend around much of the domain. The smallest field structures are
typically on finer scales than the smallest flow fields, likely partly
because we have adopted a magnetic Prandtl number Pm greater than unity.
The strongest radial fields of both polarities are generally
associated with strong downflow plumes, but only slightly weaker fields may
be found in the relatively quiescent regions between these plumes.  The
field strengths vary somewhat as a function of depth, with typical $B_r$
and $B_{\phi}$ sampled by Figure 4 declining by about a factor of two in
going from $r=0.88R$ to $r=0.24R$.

\begin{figure*}[hpt]
  \center
  \epsscale{1.0}
  \includegraphics[width=6.7in, trim= 0 0 0 0]{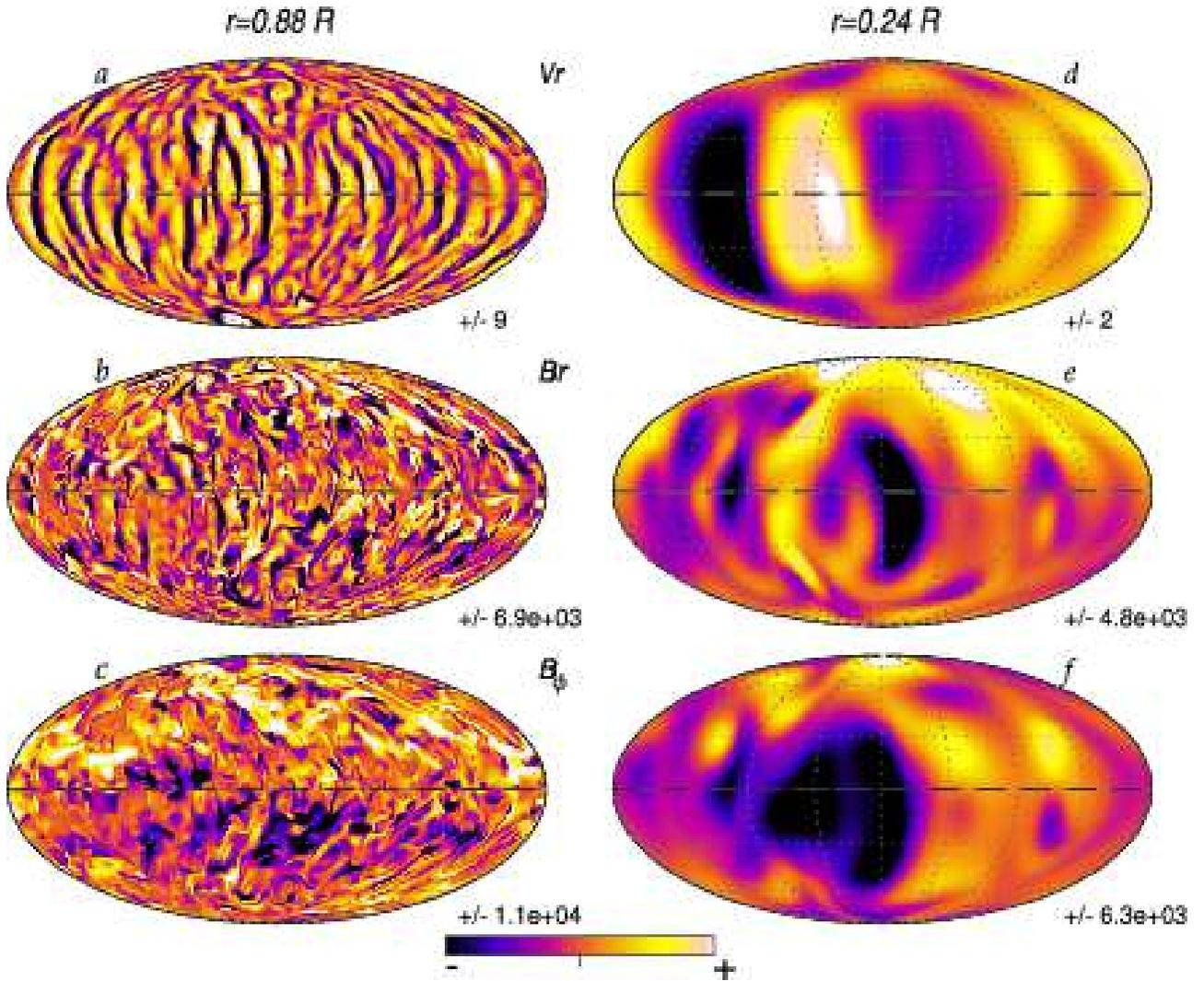}  
  \caption{\label{vbview}Global views of radial velocity $v_r$, radial
    magnetic field $B_r$, and longitudinal magnetic field $B_{\phi}$ at a
    single instant in the evolution of case Cm.  All are shown on spherical
    surfaces both deep within the star (at $r=0.24R$) and closer to the
    surface ($r=0.88R$), with red tones indicating positive polarity
    (upflows) and blue tones negative polarity (downflow).  Maxima and minima of the colormaps are
    indicated (in m s$^{-1}$ and G). }
\end{figure*}

Deeper within the star (Fig. 4 $d-f$), the magnetic fields are larger in
scale.  At these depths, the field is no longer structured on appreciably
finer scales than the flow, as revealed either by Fig 4 $d-f$ or by the
spectral analysis of \S5.2 below.  Like the convective flows, the magnetic
fields at depth are coupled to those at larger radii, with the intricate
field structures near the surface emerging from the broader network of
magnetism below.  The fields deep within the interior evolve much more
slowly than the small-scale magnetism near the surface, with some large
patterns in $B_{\phi}$ (Fig. 4$f$) persisting for thousands of days.

In the following subsections, we examine more quantitatively the strength
of the fields on both large and small spatial scales.

\subsection{Spatial Scales of the Magnetism}

One assessment of the overall field morphology is provided by decomposing
the magnetism into its azimuthal mean (the axisymmetric field) and
fluctuations about that mean.  This is a coarse measure of the size of
typical field structures: if the field is mostly on small scales, only a
small signal will survive the azimuthal averaging.  We define the
shell-averaged toroidal mean magnetic energy (TME), the fluctuating
magnetic energy (FME), and the total magnetic energy as follows:

\begin{eqnarray}
\mbox{ME}  &=&\frac{1}{8\pi}\left(B_r^2 + B_{\theta}^2 + B_{\phi}^2 \right)\,, \\  
\mbox{TME} &=&\frac{1}{8\pi}\left<B_{\phi}\right>^2 \,, \\
\mbox{FME} &=&\frac{1}{8\pi}\left((B_r-\left<B_{r}\right>)^2+(B_{\theta}-\left<B_{\theta}\right>)^2+(B_{\phi}-\left<B_{\phi}\right>)^2\right) \,,
\end{eqnarray}
recalling that the angle brackets $\left< ~ \right>$ denote a longitudinal
average.  These energy components are displayed for case Cm in Figure 5.
Although the majority of the magnetic energy is in the non-axisymmetric
component, the mean toroidal field is still considerable.  Within the bulk
of the interior, TME accounts for about 18\% of the total magnetic energy;
it is smallest near the surface, where TME $\approx$ 5\% ME, and largest
(as a fraction of ME) at depth.

\begin{figure}[hpt]
  \center
  \epsscale{1.0}
  \includegraphics[width=3.4in, trim= 10 0 10 0]{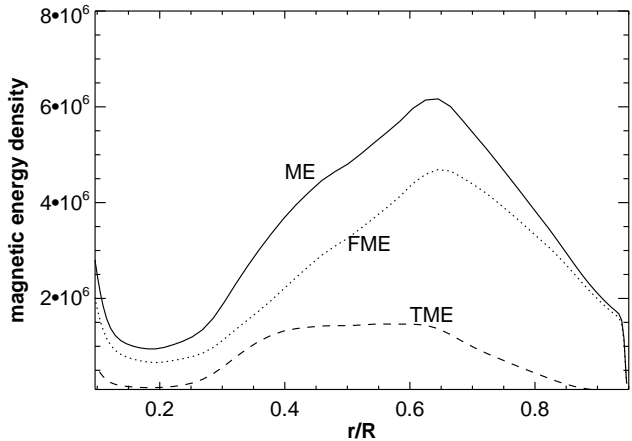}    
  \caption{\label{magbal} Magnetic energy components in case Cm as a
    function of radius.  Shown are the toroidal mean (axisymmetric)
    magnetic energy TME, the fluctuating (non-axisymmetric) magnetic energy
  FME, and their sum the total magnetic energy ME, all averaged in time and
  over spherical surfaces.}
\end{figure}

That the toroidal mean fields account for a reasonably large fraction of
the total ME is a striking result.  In prior simulations of the bulk of the
solar convective envelope, TME was typically only about 3\% (Brun et
al. 2004); in simulations including a tachocline of shear, similar TME/ME
ratios to those reported here were attained only within the stably
stratified tachocline itself (Browning et al. 2006).  Similarly, Brun et
al. (2005) found that TME/ME $\approx 0.05$ within most of the convective
cores of A-type stars, with higher values achieved only within a shear
layer at the boundary of that core.

To glean a more complete understanding of the spatial structure of the
magnetic fields and the flows that sustain them, we turn to the spatial
power spectra shown in Figure 6.  There the time-averaged $B^2$ is shown at
selected depths as a function of spherical harmonic degree $\ell$
(Fig. 6$a$), along with the velocity spectra $v^2$ for comparison
(Fig. 6$b$)  The velocity spectra broadly confirm the qualitative
descriptions of \S3.1: at large radii, the spectra peak at higher
wavenumbers (smaller spatial scales) than at small radii.  Near the
surface, the velocity amplitudes rise gradually from low $\ell$ to a peak
at about $\ell=20$, with all spherical harmonic degrees $\ell < 35$
possessing as much power as the $\ell=1$ mode.  The velocity amplitudes
fall of steeply with increasing $\ell$ beyond the peak, approximating a
power law of slope steeper than $\ell^{-2}$.  In the deeper interior, the
spectra also show a comparatively gradual rise to a maximum amplitude at a
scale $\ell_{\rm peak}$, and a steep fall-off to higher $\ell$, but the
value of $\ell_{\rm peak}$ shifts to smaller $\ell$.  At the smallest radii
sampled here, the slope of the velocity spectra becomes somewhat shallower
around $\ell=20$.  One significant caveat is that our choice
of eddy viscosities and diffusivities that are constant with depth is
somewhat arbitrary; if these coefficients were instead taken to decrease
with decreasing radius (in order to crudely represent kinetic energy
dissipation by small-scale turbulence that is constant in radius), the
contrast between the flows at depth and those near the surface would likely
not be as great as that reported here.

\begin{figure}[hpt]
  \center
\includegraphics[width=3.4in, trim= 0 0 0 0]{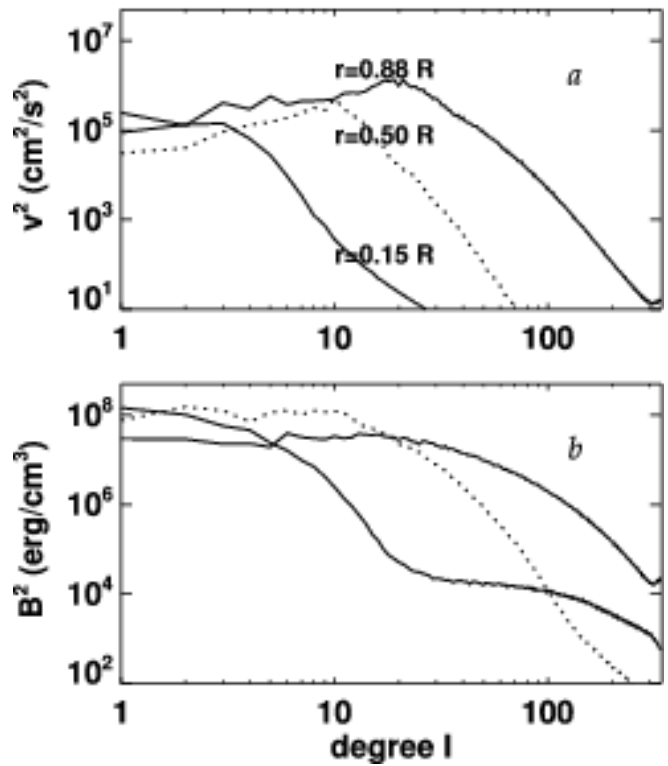}    
  \caption{\label{spectra} Time-averaged spectral distributions of ($a$)
    $v^2$ and ($b$) $B^2$ in case Cm, each sampled on three spherical surfaces
    at indicated depths.}
\end{figure}

Turning to the magnetic spectra (Fig. 6$b$), we see a somewhat different
picture.  The magnetic energy is distributed more evenly among the largest
scales: at large radii ($r=0.88R$) the spectra show a broad plateau up to
wavenumbers $\ell \approx 30$; at intermediate radii ($r=0.50R$) this
plateau extends to about $\ell=20$.  At $r=0.15R$, the magnetic energy
peaks at the largest scales ($\ell=1$) and declines continuously towards
smaller scales.  There is a break at $\ell \approx 20$ to a nearly flat
distribution of power with increasing wavenumber.  The magnetic energy in
the largest scales is actually greatest at depth, even though both the velocity
amplitudes and the total magnetic energy are smaller there.  This is partly
in keeping with the radial variation of the mean density together with the
changing scale of the flows: $\bar{\rho}$ goes from about 3.5 g
cm$^{-3}$ at $r=0.88R$ to $\bar{\rho}=86$ at 0.15R; thus the ratio of
magnetic to kinetic energy at $\ell=1$ is roughly of order unity at all
depths.

Two important points about the magnetic and velocity spectra bear
emphasizing.  One is that the distribution of magnetic energy as a function
of scale is not given simply by equipartition with the flows at each
wavenumber.  On large scales the magnetic and kinetic energies are roughly
comparable, whereas at small scales ME exceeds KE by up to a factor of 50.
A second, related point is that the spatial distributions of the magnetic
fields and flows vary appreciably with radius.  Deep within the star, the
magnetic field is dominated by its largest-scale components, while near the
surface the field is more broadly distributed in $\ell$.  These radial
variations in the ME spectra appear to be somewhat more than simple
reflections of the depth-dependent KE spectra: rather, the ratio between ME
and KE is also a function of depth, reaching its maximum at radii around
$r=0.75R$.

We caution that the power spectra presented here are likely affected by the
many simplifications and limitations of our modeling.  Both buoyancy
driving and viscous dissipation extend here over a broad range of scales,
so our simulations do not possess an extended ``inertial range'' in which
energy could simply cascade to smaller scales.  A second caveat is that our
adoption of $Pm >1$ is a likely contributor to the abundance of small-scale
magnetic energy relative to kinetic energy; this behavior is at least
expected for $\ell$ between the viscous and Ohmic diffusive scales (here
$\ell > 100$), but the amplitudes on larger scales might also be impacted.
Finally, both ME and KE show much steeper declines than expected for
homogeneous turbulence with or without magnetism; in most such models,
KE$/\ell$ and ME$/\ell$ are proportional to $\ell^{-3/2}$ or $\ell^{-5/3}$
(see, e.g., Biskamp 1993; Goldreich \& Sridhar 1995; Boldyrev 2006).  In
this respect, the spectra are qualitatively similar to those realized in
simulations of turbulent solar convection (Brun et al. 2004) or A-star core
convection (Browning et al. 2004).  However, comparison of the spectra
realized here to those expected for homogeneous, isotropic turbulence is
problematic: rotation, stratification, buoyancy driving, and the
artificially high viscous dissipation in our simulation all likely impact
the spectral energy distribution.

\subsection{Structure and Evolution of Mean Fields}

The magnetic fields realized here clearly possess both intricate small
scale structure and large-scale ordering.  We turn now to an assessment of
the large-scale mean fields, which we define to be the axisymmetric ($m=0$)
component of the magnetism.  Many different divisions into mean and
fluctuating magnetism could be employed; an axisymmetric averaging is
perhaps the simplest.  However defined, these large-scale fields hold
particular significance in dynamo theory. 

Strong axisymmetric toroidal fields are realized at many depths. These
fields are displayed for three instants in the evolution of case Cm in
Figure 7.  The magnetic fields took different times to grow at the
different depths.  Field amplification is most rapid in the outermost
regions of the star, where convection is at its most vigorous and where
typical eddy sizes are smallest.  Eventually, however, fields of comparable
strength are established in the deep interior.  The panels in Figure 7
sample $<{B}_{\phi}>$ at three times: one near $t=3500$ days on the scale of Figure
3, and the others at $t=12000$ and $14500$ days respectively.  By the time
of the first snapshot, mean fields in the outermost regions had already
grown to about 5 kG in strength , but the fields deeper down were weaker by
two orders of magnitude.  In Figure 7$b$, which shows the simulation
roughly 8500 days later, the mean fields at depth have grown to values
comparable to those at larger radii (with typical values $<B_{\phi}>
\approx 10000$ G).  We cannot reliably determine whether the fields at
depth are generated by local dynamo action there, or are instead produced
amid the more vigorous convection near the surface and then transported
downwards.  In both scenarios, strong fields are realized at depth
only on timescales reflective of the slow overturning times deep within the
star, whereas field amplification near the surface proceeds on the faster
overturning time associated with the flows there.

\begin{figure*}[hpt]
  \center
  \epsscale{1.0}
  \includegraphics[width=5.5in, trim= 0 0 0 0]{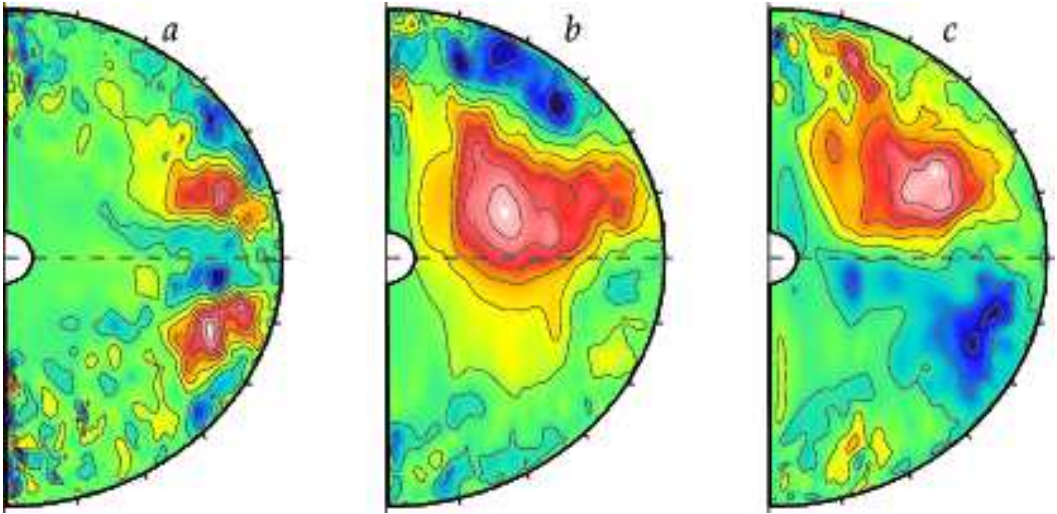}      
  \caption{\label{bevolve}Azimuthally averaged $B_{\phi}$ as contour plots
    in radius and latitude at three instants in the evolution of case Cm.
    The three renderings sample ($a$) a time prior to the saturation of the
    volume-averaged magnetic energy density, and times roughly ($b$) 8500
    and ($c$) 11000 days later.  Polarity is indicated by the colormap, with reddish tones
    positive (prograde polarity) and bluish tones negative (retrograde). }
\end{figure*}

Inspection of Figure 7 reveals that the mean fields are highly spatially
nonuniform in strength, with some regions hosting very large field
structures while others are more quiescent.  The largest $B_{\phi}$
structures extend in radius over much of the domain, and can occupy large
fractions of a hemisphere.  Comparison of Figure 7$b-c$ reveals that some
of these field structures -- in this case a prominent site of positive
polarity in the northern hemisphere -- persist over intervals of thousands
of days.  There is still substantial field evolution -- e.g. the growth of
a structure of negative polarity in the southern hemisphere (Fig. 7$c$) --
but this is most pronounced within the outer $\sim 10$\% of the
computational domain.  Once structures penetrate into the deep interior,
they appear to persist in some form for timespans more reflective of the
slow magnetic diffusion time ($\tau \sim L^2/(\pi^2 \eta) \sim 4400$ days
than of the faster convective overturning time.

The overall polarity of the fields is remarkably stable.  Over the roughly
25 years that we have evolved simulation Cm after its magnetic energy
equilibrated, the field near the surface has reversed its polarity -- which
we define as the sign of $B_r$ integrated over a surface in the northern
hemisphere (see BMT04) -- only once.  This stability is in marked contrast
to the frequent polarity reversals found in simulations of the solar
convective envelope without a tachocline (BMT04).  

\section{ESTABLISHMENT AND QUENCHING OF DIFFERENTIAL ROTATION}

Our hydrodynamical calculations (A, B, C) begin in a state of uniform
rotation.  In all of them, however, convection quickly acts to redistribute
angular momentum, ultimately establishing interior rotation profiles that
vary with radius and latitude.  The resulting differential rotation is
partly akin to that observed at the solar surface, in that the equator
rotates more rapidly than the poles; unlike the bulk of the solar
convection zone, our simulations also exhibit substantial radial angular
velocity contrasts, with the outer regions rotating more rapidly than the
interior.

This differential rotation is assessed for the hydrodynamic case C in
Figure 8.  Shown as a contour plot is the longitudinal velocity $\vph$,
averaged in time and in longitude; rapidly rotating regions are reddish,
and slower ones are bluish, all shown relative to the rotating frame.  The
rightmost panel (Fig. 8$b$) also shows the angular velocity $\hat{\Omega}$
as a function of radius along selected latitudinal cuts.  There we can see
that at the surface, the overall angular velocity contrast between the
equator and 60$\degr$ latitude is about 90 nHz, implying a $\Delta \Omega /
\Omega \approx$ 22\%.  This is comparable to the solar angular velocity
contrast $\Delta \Omega / \Omega \approx 0.25$.  As in the Sun, the angular
velocity decreases monotonically from equator to pole.  Figure 8$b$ also
reveals that $\hat{\Omega}$ generally decreases with depth, with the
equator rotating about 125 nHz faster at the top of the domain than at the
bottom.  Turning to the contour plot of $\vph$, we see that the interior
rotation profile is nearly constant on cylindrical lines parallel to the
rotation axis.  This is in keeping with the strong Taylor-Proudman
constraint felt by the flows, which are heavily influenced by rotation.
The angular velocity contrasts in radius and latitude are smaller in our
more laminar cases A and B, but the sense of the differential rotation is
the same.  We have tabulated in Table 2 the contrast from equator to
60$\degr$ for each of these simulations.

\begin{figure}[hpt]
  \center
  \epsscale{1.0}
  \includegraphics[width=3.4in, trim= 10 0 10 0]{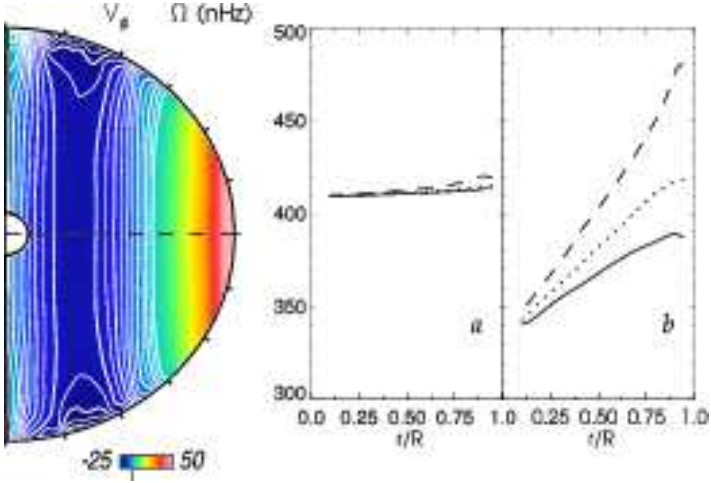}        
  \caption{\label{diffrotn} Differential rotation established in the
    hydrodynamic case C, and its quenching in the MHD simulation Cm.  Shown
  (\emph{left}) as a contour plot is the longitudinal velocity $\vph$ in
    case C, averaged in time and in longitude.  Also displayed is the
    angular velocity $\hat{\Omega}$ in ($b$) case C and ($a$) case Cm,
    shown as a function of radius along indicated latitudinal cuts.  Case
    Cm rotates essentially as a solid body.}
\end{figure}

The building of differential rotation by the rotating convective flows is
not unexpected.  As convective parcels rise and fall, they may be turned by
Coriolis forces, yielding correlations between $v_r$ and $v_{\phi}$ whose
effect is to transport angular momentum outward.  If, on the other hand,
Coriolis forces are weak (relative to buoyancy driving and pressure
forces), outward-moving flows may simply tend individually to conserve
angular momentum, implying an angular velocity that decreases with radius
(e.g., Gilman \& Foukal 1979).  The convection in our models is strongly
influenced by rotation, as quantified for instance by the Rossby number
$Ro=\tilde{u}/(L \Omega)$ or the convective Rossby number $Roc = Ra/(Ta
Pr)$.  The first of these roughly measures the strength of the Coriolis
terms in equation (3) relative to the inertial ones, while the second
estimates the influence of rotation compared to buoyancy driving.  These
are tabulated for our simulations in Table 2.  In prior studies of nonlinear
convection in rotating spherical shells (Gilman 1978, 1979; Brun \& Toomre
2002), a general finding has been that equatorial acceleration is realized
whenever $Roc$ is less than unity, with Coriolis forces therefore large.
When $Roc$ is large, conversely, the equatorial regions tend to rotate
slower than the poles.  Under strong rotational influences, angular
momentum transport by the convection tends to be radially outward and
latitudinally toward the equator (e.g., Brun \& Toomre 2002).  The analogy
in deeper spherical domains appears to be the acceleration of columns of
fluid that lie far from the rotation axis, as realized here and in the core
convection simulations of Browning et al. (2004).   Angular momentum is
globally conserved in our models, so as these regions speed up, others near
the rotation axis must slow down.

The interior rotation profiles are quite different in our calculations with
magnetism.  Intuitively, one expects that strong magnetic fields might act
like rubber bands, tying separate regions together and helping to enforce
solid body rotation.  Although this analogy is simplistic, given the
complex spatial and temporal structure of the magnetic fields realized
here, the expectation that magnetism should lessen angular velocity
contrasts turns out to be correct.  In our MHD simulations, the magnetic
fields react back strongly upon the flows, acting to strongly quench
the differential rotation.  This behavior is assessed for case Cm in Figure
8$a$, which shows the angular velocity $\hat{\Omega}$ as a function of
radius along cuts at various latitudes.  The interior is in nearly solid
body rotation; the angular velocity contrasts realized in the progenitor
case C (Fig. 8$b$) have been almost entirely eliminated.  The equator
rotates less than 2\% faster than the polar regions.  This transition towards a
uniform rotation profile is in keeping with the marked decline in DRKE
noted as the magnetic fields grew in case Cm (\S4).  In cases Cm2 and Bm,
which have lower equilibrated magnetic energy densities, the quenching of
differential rotation is somewhat less severe.  The angular velocity
contrast $\Delta \Omega / \Omega$ from the equator to 60$\degr$ is about
8\% in case Bm (compared to 14\% in the hydrodynamic case B) and about 4\%
in case Cm2.  Whether differential rotation is entirely, partially, or
minimally quenched thus seems to be a fairly sensitive function of the
magnetic energy densities realized, for ME in these three simulations
differs only by a factor of about 1.4.  Comparing the simulations here to
those of BBT05, BMT04, and Browning et al. (2006) reinforces the view that
differential rotation can persist when magnetism is weak relative to the
flows (with ME/KE less than about 30\%), is partially quenched for
intermediate-strength fields (with cyclical feedbacks between the magnetism
and differential rotation sometimes realized), and is strongly suppressed
whenever the magnetism is very strong (equipartition-strength or greater).
\emph{Our simulations show that for sufficiently turbulent flows,
such equipartition-strength magnetic fields can be realized even in fully
convective stars, and sustained once the differential rotation has been
eliminated. }

In the following section, we examine the manner in which these zonal flows
are established in the hydrodynamic simulations and quenched in the
presence of strong magnetic fields.

\subsection{Angular Momentum Transport}

The interior rotation profiles realized here result from a variety of
competing processes: angular momentum can be redistributed by Reynolds
stresses, by meridional circulations, by viscous diffusion, or by torques
and Maxwell stresses associated with the magnetic fields.  No simple
analytical tools allow us to reliably predict how each of these effects
will act, and how they will combine to shape the interior rotation
profile.  But the present simulations offer an opportunity to assess these
processes ``after the fact'': we can constrain the angular momentum
transport afforded by each and see how, together, they yield differential
rotation in the hydrodynamic cases and weaker angular velocity contrasts in the MHD
simulations.

To analyze the angular momentum transport, we turn (in the manner of
Elliott, Miesch \& Toomre 2000; BMT04) to the zonal
component of the momentum equation, expressed in conservative form and
averaged in time and in longitude:

\begin{equation}
\frac{1}{r^2} \frac{\p(r^2 {\cal F}_r)}{\p r}+\frac{1}{r \sin\theta}
\frac{\p(\sin \theta {\cal F}_{\theta})}{\p
\theta}=0,
\end{equation}
where
\begin{eqnarray}
{\cal F}_r&=&\rb r\sin\theta[-\nu r\frac{\p}{\p
r}\left(\frac{\hat{v}_{\phi}}{r}\right)+\widehat{v_{r}^{'}
v_{\phi}^{'}}+\hat{v}_r(\hat{v}_{\phi}+\Omega r\sin\theta) \nonumber \\
  &-&\frac{1}{4\pi\rb}\widehat{B_{r}^{'}
B_{\phi}^{'}}-\frac{1}{4\pi\rb}\hat{B}_r\hat{B}_{\phi}] \end{eqnarray}
and 
\begin{eqnarray}
{\cal F}_{\theta}&=&\rb r\sin\theta[-\nu
\frac{\sin\theta}{r}\frac{\p}{\p
\theta}\left(\frac{\hat{v}_{\phi}}{\sin\theta}\right)+\widehat
{v_{\theta}^{'} v_{\phi}^{'}}+\hat{v}_{\theta}(\hat{v}_{\phi}+\Omega
r\sin\theta)\nonumber \\ &-& \frac{1}{4\pi\rb}\widehat{B_{\theta}^{'}
B_{\phi}^{'}}-\frac{1}{4\pi\rb}\hat{B}_{\theta}\hat{B}_{\phi}].
\end{eqnarray}
are the mean radial and latitudinal angular momentum fluxes respectively.
The terms on the right hand side of equations (21) and (22) are, in order,
the contributions from viscous diffusion, Reynolds stresses, meridional
circulations, Maxwell stresses, and large-scale magnetic torques.  The
Reynolds stresses are associated with correlations of the fluctuating
velocity components ($v_r'$, $v_{\theta}'$, $v_{\phi}'$) that arise when
the convective structures possess organized tilts.  Similarly, Maxwell
stresses are correlations of the fluctuating magnetic field components that
correspond to the tilt and twist of magnetic structures.

To more easily analyze the various components of ${\cal F}_r$ and ${\cal
  F}_{\theta}$, we integrate over co-latitude and radius to find the net
fluxes through shells at each radius and through cones at each latitude:
\begin{eqnarray}
I_r(r)&=&\int_0^{\pi} {\cal F}_r(r,\theta) \, r^2 \sin\theta
\, d\theta \; \mbox{ , } \nonumber \\  I_{\theta}(\theta)&=&\int_{r_{bot}}^{r_{top}} {\cal
F}_{\theta}(r,\theta) \, r \sin\theta \, dr \, .
\end{eqnarray}
These are displayed for cases C and Cm in Figure 9.  We have identified
there the contributions from Reynolds stresses (denoted R), meridional
circulations (MC), viscous diffusion (V), Maxwell stresses (MS), and
large-scale magnetic torques (MT).  In constructing Figure 9, we averaged
the fluxes over about 150 days of evolution.

Turning first to the integrated radial angular momentum flux in case C
(Fig. 9$a$), we see that the Reynolds stresses act to transport angular
momentum radially outwards.  They are opposed mainly by viscous diffusion,
which transports angular momentum inwards at all radii; the flux associated
with meridional circulations plays a smaller role here, but acts to
transport angular momentum to mid-depth ($r\approx 0.66R$) from either
smaller or larger radii.  The total flux (indicated by the solid line) is
nearly zero, confirming that the rotation profile is well-equilibrated.
Note that the steady-state inward transport due to viscous diffusion
sampled here is consistent with the prevailing differential rotation in
case C: the viscous flux is negative whenever $\frac{\p}{\p r} (\vph/r)$ is
positive, as it is in case C.

\begin{figure}[hpt]
  \center
  \epsscale{1.0}
  \includegraphics[width=3.4in, trim= 0 0 0 0]{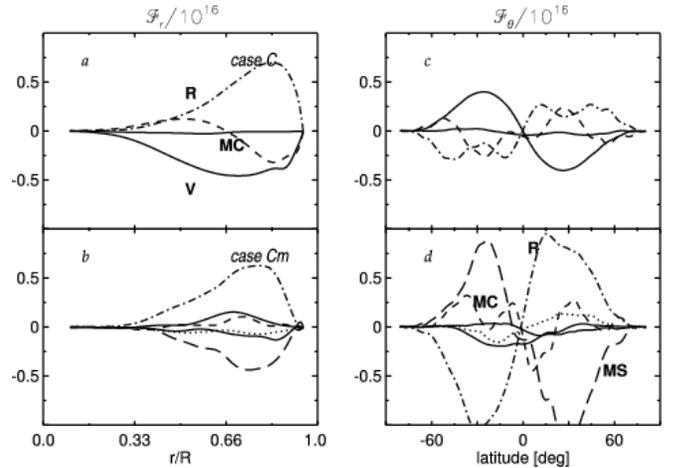}        
  \caption{\label{angmomplot} Integrated radial and latitudinal fluxes of
    angular momentum, averaged in time for case C (top panels) and case Cm
    (bottom).  Shown on left ($a$,$b$) is the radial angular momentum flux
    $I_r$.  Right panels ($c$,$d$) show the latitudinal flux $I_{\theta}$. In
    both, we have indicated the contributions from Reynolds stresses (R),
    meridional circulations (MC), Maxwell stresses (MS), viscous diffusion
    (V), and large-scale magnetic torques (MT), together with their sum
    (indicated by the solid line).  Positive quantities represent fluxes
    radially outward or directed latitudinally from north to south.}
\end{figure}

The latitudinal angular momentum flux in case C, sampled in Figure 9$c$,
conveys a similar picture.  There the Reynolds stresses act to transport
angular momentum toward the equator, since the associated flux is positive
in the northern hemisphere and negative in the southern.  Again, viscous
diffusion acts in the opposite manner, transporting angular momentum
polewards.  The meridional circulations figure more prominently than they
did in the radial balance: here they act mainly in concert with the
Reynolds stresses.

Examining the angular momentum transport in case Cm (Fig. 9$b$,$d$) reveals
that magnetic fields can play a major role in establishing the interior
rotation profile.  Figure 9$b$ shows that strong Maxwell stresses are
realized throughout much of the interior, and that they tend to transport
angular momentum inwards.  That the Reynolds and Maxwell stresses transport
angular momentum in opposite directions is understandable: the
corresponding terms in equations (20) and (21) carry opposite signs, so as
long as correlations between fluctuating magnetic field components ($\widehat{B_{r}^{'}
B_{\phi}^{'}}$) are in the same sense as the corresponding velocity
correlations ($\widehat{v_{r}^{'} v_{\phi}^{'}}$), angular momentum
  transport by Maxwell stresses will oppose that of the Reynolds stresses.
  The Reynolds stresses did not grow to compensate for the inward-directed
  flux due to Maxwell stresses; thus the region of prograde flow at large
  radii (realized in case C) was gradually slowed, yielding the nearly
  solid-body  rotation profile of case Cm.  In the equilibrated state
  sampled by Figure 9$b$, the viscous flux of angular momentum is also much
  smaller than it was in case C (Fig. 9$a$).  This is a result, rather than
  a cause, of the nearly solid-body rotation profile established in case
  Cm: the viscous transport term is proportional to $-\frac{\p}{\p r}
  (\vph/r)$, so small gradients of the angular velocity imply a vanishing
  viscous transport of angular momentum. In cases Bm and Cm2, which have
  weaker magnetic fields, the Maxwell stresses act in the same sense, but
  are weaker and so do not have as great an impact on the differential
  rotation. 

  The latitudinal transport in case Cm, sampled in Figure 9$d$, tells a
  similar story.  The dominant balance is between equatorward transport by
  the Reynolds stresses and polewward transport by Maxwell stresses.  The
  angular momentum fluxes due to viscous diffusion, meridional circulation,
  and large-scale magnetic torques are all smaller.  Somewhat surprisingly,
  the large-scale magnetic torques, though small, generally oppose the
  Maxwell stresses associated with the fluctuating fields, implying that
  correlations of the form $\hat{B}_{\theta}\hat{B}_{\phi}$ and
  $\widehat{B_{\theta}^{'} B_{\phi}^{'}}$ are of the opposite sign.

Taken together, these analyses yield some insight into how differential
rotation is established in hydrodynamic cases and quenched in MHD ones.
In the parameter regime probed here, the Reynolds stresses associated with
the turbulent convection tend to transport angular momentum radially
outward and latitudinally toward the equator.  In the hydrodynamic cases,
this transport is opposed only by viscous diffusion and, to some extent,
meridional circulations; the result is an acceleration  of regions at large
radii and low latitudes, until the steady state sampled by Figure 9$a$,$c$
(and by the contour plot of Fig. 8) is reached.  In the MHD cases, however,
the Reynolds stresses must also counteract the effect of Maxwell stresses,
which tend to transport angular momentum poleward and inward.  The
meridional circulations and large-scale magnetic torques do not adjust to
cancel out the effect of these Maxwell stresses, so the net result is a
lessening of the differential rotation.

This general picture appears robust, but we caution that some of the
\emph{detailed} features of Figure 9 -- e.g., the exact magnitude of the
viscous flux -- depend in a highly nonlinear fashion upon each other and
upon other attributes of the simulation.  The viscous flux, for instance,
depends upon the overall angular velocity gradients that are established --
but it is in part responsible for setting those angular velocity gradients,
through its competition with the Reynolds stesses, meridional circulations,
etc.  Thus viscous transport is a major player in the angular momentum
balance in case C, but a negligible one in the evolved state of case Cm: it
did not vanish because of the presence of magnetic fields, but gradually
tapered away as the Maxwell stresses lessened the angular velocity
contrast.  We have chosen for simplicity to show only the steady-state
fluxes of angular momentum in both cases; during the initial transient
phases in which the rotation profiles are established, the sense of the
fluxes -- i.e., which ones sought to speed up the equator and which to slow
it -- was generally the same as that described here.

\section{ROTATION, HELICITY, AND THE GENERATION OF FIELDS}

The magnetic fields realized here possess several striking properties.
Although substantial magnetic energy is realized on small scales, there is
also some order on the largest scales. Strong ($\sim 10$ kG) axisymmetric
toroidal fields are generated by the dynamo, and account for up to 20\% of
the magnetic energy at some sites.  Some of these strong large-scale field
structures persist for thousands of days; the overall polarity of the field
in case Cm, our longest-evolved simulation, has flipped only once in about
30 years of simulated evolution.  Furthermore, these global field
structures can be realized without the aid of stretching by differential
rotation, for the interiors of our most turbulent MHD simulations rotate
nearly as solid bodies.

These results stand in sharp contrast to those of some prior simulations of
convection in more massive stars.  In computational models of the solar
convective envelope, Brun et al. (2005) found that the toroidal mean
magnetic energy was typically less than 5\% of the total ME; the polarity
of the mean field typically reversed at chaotic intervals of less than 600
days.  In simulations that also included the tachocline below (Browning et
al. 2006), higher values of TME/ME were achieved only amid the strong shear
of the tachocline itself; the polarity evolution of the fields was
stabilized by the presence of strong mean fields in the radiative layer.
In neither of these sets of simulations did the magnetic energy become
strong enough to quench the differential rotation entirely.  Similarly,
models of dynamo action in the convective cores of A-type stars (Brun et
al. 2005) indicated that strong large-scale fields were mostly realized in
the shearing layer near the core-envelope boundary.  Those simulations also
exhibited a rich variety of interactions between the magnetism and the
differential rotation: angular velocity contrasts were almost entirely
eliminated in rapidly rotating cases, while slower rotators showed cyclical
feedbacks between differential rotation and magnetism.

Some guidance in interpreting these results may be afforded by findings
from mean field theories (MFTs) of the dynamo process.  In many such
theories, a key role is played by the kinetic helicity, ${\bf
  v}\cdot\nab\times{\bf v}$, which is related to the twisting and winding
of the convective flows (see Moffatt and Tsinober 1992).  Under a number of
simplifying assumptions, it can be shown that the ``$\alpha$-effect'' of
traditional MFT is proportional to the kinetic helicity of the turbulence,
implying that more helical flows might build strong large scale flows
(e.g., Parker 1955; Steenbeck et al. 1966; Moffatt 1978; Moffatt \& Proctor 1982; Brandenburg \& Subramanian
2005).  This general expectation has been partly born out by simulations of
dynamo action in forced helical turbulence and by numerical calculations
using turbulent closure schemes (e.g., Blackman 2003; Pouquet \& Patterson
1978).  In such modeling, the spectrum of the magnetic fields realized by
dynamo action changed as the kinetic helicity was varied, with more
large-scale field generated when the helicity was increased.  A common
expectation is thus that flows without helicity can
act as dynamos, but the fields are typically on the scale of the turbulent
eddies (e.g., Brandenburg \& Subramanian 2005).  More recent
numerical modeling has suggested, however, that the linkages between kinetic helicity
and large-scale magnetic fields may not be so clear -- e.g., Courvoisier,
Hughes \& Tobias (2006) found no relation between the $\alpha$-effect and
kinetic helicity in models of a variety of chaotic flows.  

The convective flows in our simulation are strongly influenced by rotation.
The local Rossby number $Ro_w = \nab \times {\bf v}/2 \Omega$ (which
compares the fluid vorticity to the ``planetary'' vorticity $2\Omega$) is
less than 0.1 throughout most of the interior; its radial variation for
case Cm is shown in Figure 10.  Also shown for comparison there is the
globally averaged value of $Ro_w \approx 0.86$ for the penetrative solar
dynamo calculations of Browning et al. (2006) (calculated within the bulk
of the convection zone).  The local Rossby number here is significantly
less than in the solar simulation, even though the angular velocity in case
Cm is equal to the solar rate of $\Omega = 2.6 \times 10^{-6}$
s$^{-1}$. Likewise, the global estimates of $Ro$ and $Roc$ given in Table 2
are significantly lower than those calculated for the solar simulations of
BMT04 or Browning et al. (2006).  The influence of rotation is stronger
here for the same rotation rate because the stellar luminosity is much
lower, so typical convective velocities are slower than in the solar case,
and the convective overturning time is longer.  Flows that take more than
one rotation period to overturn can be strongly affected by Coriolis
forces, whereas those that overturn faster cannot; here only the rapid,
small-scale flows near the surface have overturning times that approach the
rotation period.  Note that the rotation rate adopted here is lower than
that observed for most mid-late M-stars, so nearly all such stars are
probably strongly influenced by rotation.

\begin{figure}[hpt]
  \center
  \epsscale{1.0}
  \includegraphics[width=3.4in, trim= 0 0 0 0]{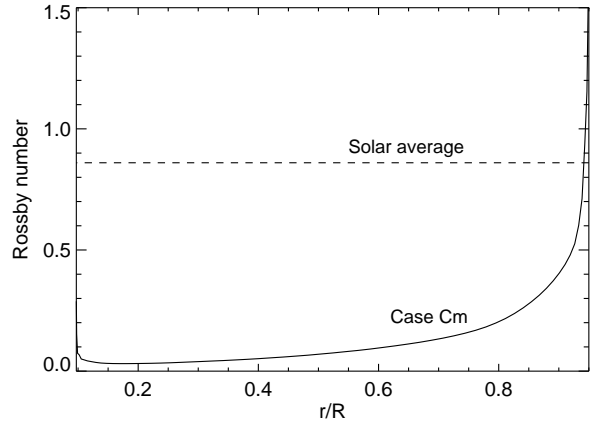}          
  \caption{\label{kinhel} Local Rossby number $Ro = (\nabla \cross u)/2
    \Omega$, averaged in time and on spherical surfaces.  Also shown is the
  average value of $Ro \approx 0.86$ within the solar convection zone
  simulations of Browning et al. (2006).  The lower Rossby numbers realized
  here indicate a stronger rotational influence.}
\end{figure}

The general trend that emerges from comparing our simulations to prior ones
is that a stronger rotational influence, and hence lower $Ro$, implies both
higher magnetic energy densities relative to kinetic, and magnetic fields
of increasingly large spatial scale.  When $Ro$ is close to unity, as in
the Sun (Browning et al. 2006; Brun et al. 2005), the magnetic energy
generally appears to be small enough that differential rotation can readily
persist.  Under somewhat stronger rotational influences, as realized in
some of the A-star core dynamos of Brun et al. (2005), the differential
rotation and magnetism may feed back upon each other, possibly yielding
cyclical waxing and waning of field strength.  At the still stronger
rotational influences sampled here and in the more rapidly rotating cases
of Brun et al. (2005), the helical convective flows are able to build
magnetism of equipartition strength without the aid of differential
rotation; angular velocity contrasts realized in hydrodynamic cases are
greatly reduced by the strong magnetic fields.

Simulations of the geodynamo also lend support to the idea that stronger
rotational influence can lead to magnetic fields on larger spatial scales.
Christensen \& Aubert (2006) found that lower $Ro$ led to magnetic fields
with a larger ``dipole fraction,'' defined as the power in the $\ell=1$
mode divided by that in modes $\ell=1-12$.  They and others have suggested
that predominantly dipolar fields are realized when inertial forces are
small relative to Coriolis forces (see also Sreenivasan \& Jones 2006;
Olson \& Christensen 2006).  The transition between dipolar and multipolar
fields in their models occurred at a ``modified Rossby number'' $Ro_l
\approx 0.1$, where $Ro_l = Ro\frac{\ell_u}{\pi}$, with $\ell_u$ the mean
spherical harmonic degree of the flows.  Constructing a local $Ro_l$ as a
function of depth in the simulations here (using the power spectra in
\S5.2) suggests that our models may be on the cusp of entering into the
predominantly dipolar regime identified by Christensen \& Aubert (2006).
At most depths, $Ro_l$ in our simulations is still somewhat greater than
0.1, and the dipole fraction is low; deep in the interior, however, $Ro_l
\approx 0.1$, and the local dipole fraction rises to about 30\%.  These
results are suggestive of the role that rotation may play in setting the
field geometry, though clearly much more work is needed to clarify how
stratification, dynamo supercriticality, and other effects might likewise
enter into the field strength and morphology.

Quantifying the connection between rotational influence (as measured by
$Ro$) and kinetic helicity, which figures so prominently in MFT, is a
complex task.  The two are clearly related, for it is partly the overall
rotation that imparts a global ordering to the helicity: downflows in the
northern hemisphere tend to contract, and because of Coriolis forces rotate
counter-clockwise, implying anti-cyclonic vorticity (e.g., Miesch et
al. 2000).  Thus on average the kinetic helicity is negative in the
northern hemisphere and positive in the southern hemisphere.  A naive
expectation might therefore be that as Coriolis forces become increasingly
important, this process would lead to stronger net helicity.  Indeed, in
simple models, the average kinetic helicity is often taken to be
proportional to the overall angular velocity (e.g., Durney, Mihalas \&
Robinson 1981; Noyes et al. 1984), reflecting the ability of Coriolis forces to
twist convective parcels as they rise or descend.  Under such an
approximation, the ``dynamo number'' of MFT may then be proportional to the
angular velocity squared (e.g., Noyes et al. 1984). Such direct connections
between helicity and rotation rate are not realized in our simulations.
Both the azimuthally-averaged kinetic helicity and its rms values are
smaller here than in, for instance, the solar calculations of Browning et
al. (2006), even though Ro is substantially lower.  This probably reflects
several key differences between the flows here and in the solar
simulations: the velocities here are lower, and the stratification weaker
throughout most of the interior, leading to less asymmetry between upflows
and downflows, and hence less preference for one sign of helicity.
Furthermore, the convection here shows some tendency to align in rolls
parallel to the rotation axis, reflecting the strong Taylor-Proudman
constraint; in such rolls, the velocity is mostly perpendicular to the
rotation axis, but the vorticity is mostly parallel to it, resulting in a
small average kinetic helicity (e.g., Knobloch, Rosner \& Weiss 1981).  A
further complication is that recent numerical modeling and asymptotic
analysis of {\sl unstratified} turbulence has shown that a preference for
one sign of vorticity, and hence a high net helicity, is established at
moderate rotation rates with $Ro \approx 0.1 \rightarrow 1$, but eliminated
as rotation becomes even more rapid (Sprague et al. 2006).  This results in
little net helicity in the most rapid rotators.  Whether such analysis is
relevant to the stratified flows here is not clear at this stage, but bears
further study.

Drawing the further connection between kinetic helicity and the generation
of large-scale magnetic fields in our simulations appears to be even more
difficult.  Like Livermore, Hughes \& Tobias (2007), we have seen no clear
linkages between the magnitude or power spectrum of the helicity and that
of the magnetic field.  Indeed, the power spectrum of this quantity in the
present simulations is qualitatively similar to one constructed for the
convective envelope in the solar simulations of Browning et al. (2006).  As
noted above, the magnitude of the kinetic helicity is smaller here than in
those simulations or the ones of BBT05, yet the axisymmetric magnetic field
is stronger. More sophisticated theories relating the growth of fields to
the kinetic helicity appear required; alternatively, it may be that the
dynamo action at some scales could be partly ``quenched'' by the growth of
current helicity at those scales, as suggested in some variants of MFT
(see, e.g., Blackman 2003b).  Much further work will be required to assess
whether such theories or variations thereof can accurately describe the
growth of fields in the complex flows here.

We close this section with brief comments on the strength and temporal
variability of the fields realized here.  That mean toroidal fields of
order 10 kG strength can be maintained in a fully convective star for long
periods of time might come as a surprise: in the solar convective envelope,
simple estimates suggest that fields of that strength would quickly rise
due to magnetic buoyancy (Parker 1975). Indeed, the rapid timescale for
such rise was a major factor in identifying the stable layer as the likely
seat of the global solar dynamo.  Here, however, the limits on field
strength imposed by magnetic buoyancy are less severe.  A crude estimate of
the upward velocity for a buoyant flux tube in an unstably stratified layer
is
\begin{equation}  
  u \sim v_a \left( \frac{\pi a}{C_d \Lambda}\right),
\end{equation}
with $\sim v_a = B/\sqrt(4 \pi \rho)$ the Alfven speed, $C_d$ the
aerodynamic drag coefficient, $a$ the tube radius, and $\Lambda$ the
pressure scale height (Parker 1975).  For constant $\frac{a}{\Lambda}$ and
$C_d$, this estimate implies that the characteristic timescale for fields
to rise from the center of an M-dwarf to its surface is about a factor of
ten longer than the time needed for fields to rise through the solar
convective envelope.  The main difference is that the density in the M-star is
substantially greater, so the Alfven speed is lower and the rise time
greater.  Finally, we note that the long temporal stability of the field
polarity here is also striking, but may partly reflect the lower typical
convective velocities realized here.  If the field is modeled as a
collection of overlapping dipole moments corresponding to typical
convective eddies, and these individual eddies are uncorrelated, then the
overall magnetic dipole should evolve on a random-walk timescale, $\tau_B
\sim R^2/\nu_{t}$, with $\nu_t \sim v_c l_{\rm ed}$ (Thompson \& Duncan
1993).  This estimate would imply $\tau_B \sim 6$ years for motions near the
surface.  The field actually appears to evolve on somewhat slower
timescales that are more characteristic of the flows deep in the interior.
A possibly analogous effect was found in Browning et al. (2006), where the
presence of organized mean fields deep in the interior served to largely
stabilize the sense of fields produced in the convection zone.  This is a
cautionary tale: the field dynamics near the stellar surface may well
reflect couplings with an interior field that is hidden from view.

\section{CONCLUSIONS AND PERSPECTIVES}

We have presented global 3-D MHD simulations of the interiors of fully
convective M-stars rotating at the solar angular velocity.  These nonlinear
simulations attempt to model a 0.3 solar-mass star with reasonable
fidelity: the stratifications of density and temperature are consistent
with those of a 1-D stellar model, as are the luminosity and rotation rate.
Although many great simplifications have been made, we believe that several
of the conclusions of our work may prove to be robust.  We reiterate these
principal findings below, and comment briefly on the uncertainties
associated with each.  We also compare our work to the simulations of
DSB06, and to the limited observational constraints presently available.

The convection realized in these simulations is characterized by
small-scale, intermittent flows near the stellar surface, and by weaker,
large-scale flows in the deep interior.  The radial variation in the size
of typical convective eddies is driven mainly by the strong density
stratification (with $\rho_{bot} \approx 100 \rho_{top}$).  The
weakening of convective velocities at depth arises partly because the
luminosity $L_c$ that must be carried by the convection -- essentially the
difference between the total luminosity $L_t$ and the radiative luminosity
$L_r$ -- is quite small there, even though the star is everywhere unstably
stratified.  Thus in the cores of our model stars, convection is fairly weak
and radiation actually still transports much of the energy.  The relatively modest energy transport
afforded by convection in the deep interior does not result directly from
convective suppression by magnetic fields: convection is weak at depth even
in hydrodynamic calculations.

The flows act as a magnetic dynamo, amplifying a small seed field by orders
of magnitude and sustaining it against Ohmic decay.  The equilibrated
magnetic energy density in our highest resolution, most turbulent
simulation Cm is roughly 120\% \ of the kinetic energy density relative to
the rotating frame.  The resulting magnetic fields possess structure on a
wide variety of spatial scales; the typical size of field structures is
largest in the deep interior, and smaller near the surface.  Strikingly,
the magnetic field possesses a strong axisymmetric mean component, with the
toroidal mean energy accounting for up to 20\% \_ of the total magnetic
energy.  Such prominent large-scale mean fields have not been realized in
simulations of the solar convection zone (Brun, Miesch \& Toomre 2004),
except within a stably stratified tachocline of shear (Browning et
al. 2006).  The mean fields realized here also possess remarkably stable
polarities: only one reversal of the overall polarity was realized in the
roughly 25-year evolution of case Cm.  This result, too, stands in contrast
to prior simulations of solar-like convection in spherical shells (Brun et
al. 2004), in which the overall polarity tended to flip at irregular
intervals of less than 600 days.

Differential rotation is established by the convection in hydrodynamic
cases, but reduced in MHD simulations.  In our most turbulent case Cm,
which has the strongest dynamo-generated magnetic fields, the differential
rotation of the hydrodynamic progenitor is almost entirely eliminated.
This occurs because of strong Maxwell stresses, which tend to oppose the
equatorward transport of angular momentum by Reynolds stresses.  In the
non-magnetic cases, the differential rotation is solar-like at the surface,
with a fast equator and slow poles; the interior rotation profile is
largely constant on cylinders, in accord with the strong Taylor-Proudman
constraint.  The angular velocity contrasts in those hydrodynamic cases are
fairly modest, in keeping with simulations of rapidly rotating solar-like
stars (Brown et al. 2007); it is possible that this reflects non-magnetic
mechanisms for quenching zonal flows in rapidly rotating systems, as
discussed in mean-field models of angular momentum transport (e.g.,
Kitchatinov \& Rudiger 1995).  No cyclical feedbacks between the
differential rotation and the magnetic field -- noted in prior simulations
of core convection in A-type stars at certain rotation rates (BBT05) -- are
seen in case Cm.  Rather, equipartition-strength fields are sustained even
in the absence of any differential rotation.  More interplay between the
differential rotation and magnetic fields is realized in the two cases Bm
and Cm2, in which the magnetism is weaker relative to the kinetic energy
and the differential rotation is somewhat more persistent.  The magnetic
fields also impact the convective (non-axisymmetric) flows, though much less
drastically; this weakening of the convection may be somewhat akin to that
explored by Chabrier, Gallardo, \& Baraffe (2007) in the context of
mixing-length theory.

We have argued in \S7 that several key attributes of these simulations --
the strength of the magnetic fields realized, their strong large-scale
axisymmetric components, and the quenching of differential rotation -- may
depend crucially on the influence of rotation.  Although the
simulations here rotate at the solar angular velocity, the rotational
influence upon the convective flows is far stronger than in simulations of
the solar convective envelope (e.g., Browning et al. 2006) or the cores of
A-type stars rotating at the same angular velocity (BBT05).  Here, the very
slow convective flows imply Rossby numbers significantly smaller than in
more luminous stars at the same rotation rate.  A comparison of our
simulations here to others at varying Rossby number (BBT05; Browning et
al. 2006) suggests that as the rotational influence becomes stronger, the
magnetic energy density grows larger relative to the kinetic energy
density.  When the ratio ME/KE is small ($<$ 30\%), strong differential
rotation can persist; as ME grows larger (with increasing rotational
influence), a regime may be reached in which the differential rotation and
ME each undergo cyclical waxing and waning.  Finally, in stars where
rotation is still more important -- as achieved in the 4-$\Omega_{\sun}$
cases of BBT05 and in our simulations Cm and Cm2 here -- ME can exceed KE without the
aid of differential rotation, and any persistent angular velocity contrasts
are strongly quenched.  These conclusions are also tentatively born out by
preliminary simulations of dynamo action in more rapidly rotating
solar-like stars (Brown et al. 2007, in prep).

Thus, we think it likely that had the simulations here been rotating at
only a quarter of the solar angular velocity (to yield Rossby numbers more
in accord with the 1-$\Omega_{\sun}$ simulations of BBT05), cyclical
feedbacks of ME upon DRKE would have been obtained even in our most
turbulent cases.  At still slower rotation rates, the differential rotation
established in hydrodynamic cases would likely have persisted in the
presence of the weaker sustained magnetism.  But the simulations here
already correspond to rotational velocities below what can presently be
measured: for a star with a radius of $2 \times 10^{10}$ cm, the solar
angular velocity of $\Omega_{\sun}=2.6 \times 10^{-6}$ s$^{-1}$ implies
$v_{\rm rot} \approx 0.5$ km s$^{-1}$, below current detection limits of
vsini $\approx 2$ km s$^{-1}$ (e.g., Delfosse et al. 1998).  From the point
of view of observations, even the models presented here would be
``non-rotating.''

Like all those who would simulate stellar convection, we have made
major simplifications in our modeling.  Of these, our use of effective eddy
viscosities and diffusivities -- which are vastly greater than their
microscopic counterparts in actual stars -- is arguably the most severe.
Other potentially important simplifications include our omission of the
inner and outer few percent of the stellar interior (our computational
domain extends from 0.08 to 0.96R), our adoption of a perfect gas equation
of state, and the limited duration of the simulations compared to the very
long thermal relaxation timescales in actual stars.  It is difficult to
estimate the impacts each of these may have, but we are encouraged by the
reasonable success that similar simulations of the solar interior have
enjoyed in matching the detailed observational constraints provided by
helioseismology (Miesch et al. 2006; Browning et al. 2006).

Probably the greatest uncertainty associated with these simplifications is
the extent to which changes in other simulation parameters -- e.g., the
magnetic Reynolds and Prandtl numbers -- could mimic the effects of
rotation upon magnetic field strength and differential rotation.  Both the
strength and morphology of the field are likely sensitive to these
parameters at some level, as indicated by the modest differences in ME and
DRKE between our 3 cases Cm, Cm2, and Bm.  Our simulations were conducted
in an Rm, Pm regime probed frequently by prior simulations, and yet they
exhibit quite distinct behavior.  The closest analogues in our own prior
work are the most rapidly rotating cases of BBT05; this fact, and a
comparison to cases at other Rossby numbers, is a partial motivation for
our suggestion that the Rossby number is a dominant control parameter in
determining the strength and geometry of the magnetic fields.  Future work
that more thoroughly probes the parameter space of rotation rate, Rm, and
Pm will be needed in order to put this tentative suggestion on firmer
ground.

Our results partly agree with those of DSB06, in that we find that strong,
equipartition-strength magnetic fields can be generated at certain rotation
rates.  Like them, we also find that the fields have substantial
axisymmetric mean components, and that they possess structure on both large
and small spatial scales.  The most significant differences between our
findings and theirs concern the differential rotation established in the
interior: they find that hydrodynamic simulations establish an anti-solar
differential rotation, and that this differential rotation persists even
when strong magnetism is present.  We find solar-like differential rotation
in our hydrodynamical models, and nearly solid-body rotation in our most
turbulent MHD simulations.  Although the origins of this discrepancy in our
results are not certain, it may be partly caused by the differing strengths
of rotation and turbulence in our simulations compared to those of DSB06.
Although the rotational velocities adopted in DSB06 are far greater than in
actual stars, so are the convective velocities (because the stellar
luminosity was greatly enhanced).  It is difficult to say how this
rescaling affects the delicate balance of convection, rotation, and
magnetism, but it is reasonable to suggest that the overall rotational
influence in most of their simulations is somewhat smaller than in ours.  A
crude estimate of the rotational influence is given by $Ro \approx (u_{\rm
  rms}/R)/2\Omega$; for most of their simulations, Ro is greater than in
ours, indicating that rotation is weaker relative to inertia. Only for
their two most rapidly rotating cases does Ro drop below the values
reported for our cases. They note a trend towards increasing ME/KE with
increasing rotation rate in their simulations; they also note that the
overall angular velocity contrast is reduced as the rotation rate is
increased (in their magnetic cases).  Thus we suspect that for somewhat
more rapid rotation, they too would find even stronger quenching of the
differential rotation, perhaps yielding solid-body rotation like that
realized in our case Cm.  Alternatively, at fixed rotation rate, more
turbulent flow (and higher $Rm$ and $Re$) may lead to stronger magnetic
energy densitites (as evinced by a comparison of case Bm to case Cm), and
hence to stronger quenching of the differential rotation.  Indeed, our case
Cm has a somewhat higher $Re$ (and significantly higher $Rm$) than the
simulations of DSB06 that have comparable rotation rates; our more laminar
case Bm possesses magnetic energy densities and angular velocity contrasts
somewhat more akin to those of DSB06.  Other differences between our
simulations and those of DSB06 also impact the results in subtler ways.
For instance, the morphology of the convective flows and their variation
with radius are strongly affected by the density stratification; our
overall density contrast of $\approx 100$ is consistent with a 1--D stellar
model, whereas theirs is a factor of 20 smaller, so naturally the developed
flow patterns in our simulations differ somewhat.  Other differences --
e.g., the boundary conditions adopted on the flow fields -- may also impact
the results at some level.

If we are correct in suggesting that rotational influence is the dominant
control parameter in setting the magnetic energy, the field morphology, and
(indirectly, through the feedback of Maxwell stresses) the differential
rotation, then some straightforward observational predictions follow.
Rapidly rotating M-stars should generally show strong magnetic fields and
little or no differential rotation.  Less rapidly rotating ones may show
cyclical magnetic energy and differential rotation, while the slowest
rotators are most likely to harbor persistent angular velocity contrasts,
accompanied by somewhat weaker average magnetic energies.  Again
extrapolating from our limited probing of parameter space, the axisymmetric
component of the field should account for a greater fraction of the
magnetic energy in progressively more rapid rotators.  Although the
observational constraints on magnetic fields and differential rotation in
fully convective M-stars are still scarce, it appears that these
predictions are at least consistent with what has so far been observed.
Donati et al. (2006) reported that v374 Peg had a strong, mostly
axisymmetric magnetic field, with no evident differential rotation.  Their
target star was very rapidly rotating, in keeping with our suggestion that
differential rotation should be strongly quenched in such stars.  At still
lower masses, Reiners \& Basri (2007) have found that magnetic activity is
detectable even in some stars that are not detectably rotating; such stars
may conceivably have low enough convective velocities that even rotation
rates below the observational detection limit could imply a reasonably
strong rotational influence and hence lead to vigorous dynamo action.  Much
more work will be required to elucidate the full role of rotation in such
stars, and to determine whether other effects ignored in our modeling --
e.g., degeneracy and decreasing surface conductivity -- also play roles in
the dynamo process.

\acknowledgements
It is a pleasure to thank Gibor Basri, Juri Toomre, A. Sacha Brun, Andrew
West, Mark Miesch, and Benjamin Brown for helpful discussions and/or
comments on this manuscript.  We also gratefully acknowledge Isabelle
Baraffe and Gilles Chabrier for supplying the 1--D stellar model used here
for initial conditions.  This work was supported by an NSF Astronomy \&
Astrophysics postdoctoral fellowship (AST 05-02413).  The simulations were
carried out with NASA support of Project Columbia, and NSF PACI support of
NCSA, SDSC, and PSC.

\clearpage

\end{document}